\documentclass[a4paper,11pt]{article}
\usepackage{jhepmod} 
\usepackage{lineno}
\usepackage{lmodern}

\usepackage{amsmath,amssymb,amsbsy,amstext,amsthm,amsfonts,mathrsfs,bm,ulem,hyperref}
\usepackage{xcolor}
\usepackage[framemethod=default]{mdframed}
\graphicspath{{Fig/}}

\newmdenv[backgroundcolor=gray!10,%
skipabove=5pt,%
skipbelow=7pt,%
leftmargin=0pt,%
rightmargin=0pt,%
innertopmargin=-6pt,%
innerbottommargin=5pt,%
innerleftmargin=5pt,%
innerrightmargin=5pt,%
splittopskip=0pt,%
splitbottomskip=0pt,%
linewidth=0pt,%
nobreak=true]%
{keyeqn}

\title{Graviton stimulated emission in \\ squeezed vacuum states}

\author[1]{Atsuhisa Ota}
\author[2]{and Yuhang Zhu}
\affiliation[1]{Department of Physics and Chongqing Key Laboratory for Strongly Coupled Physics, \\
Chongqing University, Chongqing 401331, People's Republic of China}
\affiliation[2]{Cosmology, Gravity and Astroparticle Physics Group,\\
        Center for Theoretical Physics of the Universe,\\
        Institute for Basic Science, Daejeon 34126, Korea}
\emailAdd{aota@cqu.edu.cn}
\emailAdd{yhzhu@ibs.re.kr}
\abstract{
We study the dynamics of gravitons in a squeezed vacuum state under a thermal radiation background. Unlike traditional treatments that rely on the Boltzmann equation, we employ the Heisenberg equation and average it over general quantum states. In contrast to the usual Boltzmann-based descriptions, our approach captures the subtleties arising from quantum coherence in different number eigenstates, which is essential for soft graviton modes in the squeezed vacuum state. Our new method successfully reproduces the previous one-loop results within the in-in formalism when the expansion parameter is small and deviates significantly as the parameter increases, indicating that our results extend beyond the one-loop in-in formalism. We examine the implications of graviton emission effects stimulated by quantum coherence in both flat and expanding backgrounds. In the flat background, it is found that backreaction of radiation on the spacetime dynamics is crucial for significant stimulated emission. In the expanding background, to avoid the subtleties associated with superhorizon modes, we investigate the effect of emission within the horizon immediately after reheating and find a significant effect. We also examined the IR graviton evolution from a symmetry perspective and propose a regularization prescription to eliminate the secular growth problem.
}

\begin{document}
\maketitle
\flushbottom

\section{Introduction}
\label{sec:intro}
Gravity is omnipresent, intricately coupled across all scales of known physics, yet it is the weakest in coupling strength among all fundamental forces, making it exceedingly difficult to capture its presence. While the elegant framework of General Relativity was established already a century ago, the direct detection of gravitational waves was only recently confirmed with extremely sensitive experiments \cite{LIGOScientific:2016aoc,LIGOScientific:2017vwq}. In inflationary cosmology, although the curvature fluctuations predicted by inflation are already observed in the anisotropies of the cosmic microwave background, primordial gravitational waves remain elusive. Despite the fact that inflation might be the highest energy scale accessible to us, due to Planck mass suppression, the strength of such signals is extremely small, and now the tensor-to-scalar ratio is constrained to $r \lessapprox \mathcal{O}(0.01)$~\cite{Planck:2018vyg}. To capture more signals of gravitational waves, we either keep improving the sensitivity of experiments or look for more dramatic sources, such as the merger of more massive compact objects, large scalar perturbations acting as secondary sources \cite{Ananda:2006af,Baumann:2007zm,Saito:2008jc,Kohri:2018awv,Domenech:2021ztg}, and so on.

Under the assumption of general relativity, the coupling between matter and gravitons is ubiquitous. The extreme weakness of gravitational interactions stems from the smallness of the gravitational coupling constant. As a consequence, scattering processes in vacuum are exceedingly rare, rendering gravitational interactions difficult to detect. A natural way to compensate for the small interaction strength is to increase the number of target particles that interact with gravitons. For example, the relatively large gravitational waves can be generated if the source particles are produced in large numbers through certain instabilities \cite{Barnaby:2011qe,Cook:2011hg,Anber:2012du,Salehian:2020dsf,Wang:2020uic,Niu:2022quw,An:2025mdb}. Another natural possibility is to consider the high temperature thermal bath, since particle occupation numbers increase with temperature. In an extreme scenario where the temperature of the thermal bath is high enough to drive spacetime evolution, the universe enters a radiation-dominated phase. In this regime, the Friedmann equation then relates the temperature to the Hubble parameter $H$ as $T \sim \sqrt{M_{\rm pl} H}$. Nevertheless, as demonstrated in the previous work \cite{Ota:2023iyh}, these thermal induced gravitational waves may still be too weak, since their amplitude remains suppressed by $H/M_{\rm pl}$. Is there an alternative mechanism to further enhance the amplitude of gravitational waves? Motivated by light amplification via stimulated emission in atomic systems, we will consider the possibility of graviton stimulated emission from a thermal bath \cite{Ota:2024idm}.

To fully understand the mechanism underlying the possibility of graviton stimulated emission, it is essential to carefully specify the quantum states of the gravitons. If we assume that gravitons are in a mixed state expanded in terms of Hamiltonian eigenstates as adopted in many previous studies, then the Heisenberg equation of quantum operators can be reduced to a Boltzmann equation for the phase space distribution. This method, sometimes referred to as the quantum Boltzmann equation, was developed in the study of neutrino mixing~\cite{Raffelt:1992uj,Sigl:1993ctk} and has also been applied to the polarization of the cosmic microwave background~\cite{Kosowsky:1994cy} (for recent developments, see also \cite{Alexander:2008fp,Bavarsad:2009hm,Mohammadi:2013dea,Fidler:2017pkg,Bartolo:2018igk,Bartolo:2019eac, Manshouri:2020avm,Zarei:2021dpb,Sharifian:2025olk}).
Such an ansatz for a quantum state permits counting graviton numbers classically and applying a kinematic description. In such a framework, the phase space distributions of each polarization mode fully characterize statistical graviton states. However, this is not necessarily the case for general quantum states of gravitons. For instance, the distribution of primordial gravitons generated during inflation reside in a two-mode squeezed vacuum state~\cite{Albrecht:1992kf}, which is not solely determined by graviton number, as the coherence of different number eigenstates is nonvanishing.  As in Ref.~\cite{Ota:2024idm}, it was suggested that squeezed vacuum states of gravitons could stimulate  graviton emission, potentially enhancing the initial graviton population~\footnote{As another interesting direction is graviton production in the thermal background in hydrodynamical scale~\cite{Ghiglieri:2015nfa,Ghiglieri:2020mhm,Ghiglieri:2022rfp}, while we focus on the quantum effect on IR relic gravitons in this paper.}. This setup is incompatible with the usual assumptions in kinetic theory and requires the inclusion of additional information.

To be more specific, rather than focusing solely on the evolution of the number operator as in the conventional Boltzmann approach, we will introduce a new operator $L_{\mathbf{k}}$ which encodes the coherence between different number eigenstates. We will show that the dynamics of the number operator and this new operator are coupled and governed by a set of first-order differential equations. We can thus evaluate the stimulated emission of gravitons in squeezed states by solving this set of time-evolution differential equations. This new approach is also different from the earlier work \cite{Ota:2023iyh,Ota:2024idm}, which is based on perturbation theory and used the in-in formalism. Instead of dealing with differential equations, that method involves multiple nested time integrals, making it difficult to go beyond one-loop calculations.

The above effect, termed \textit{cosmological stimulated emission}, presents two key challenges.
First, the emission rate is proportional not to the number density but to the radiation pressure.
As a result, the usual Planck suppression of the emission rate is canceled, potentially causing a breakdown of perturbative analysis.
Second, the additional mass dimension introduced by radiation pressure is compensated by graviton momentum, implying that infrared (IR) gravitons are more sensitive to the process.
Such an IR sensitivity is common for the stimulated emission for photons~\cite{Weinberg:2015QM}. 
In lasers, the finite gap of atomic electron eigenstates limits the maximum possible IR photon wavelength~.
However, in a cosmological situation, there is no such constraint since the radiation momenta can form arbitrary squeezed triangle configurations in momentum space, which results in a secular growth of the IR graviton power spectrum~\cite{Ota:2023iyh}, that has also been confirmed recently by \cite{Frob:2025sfq}.

In this paper, to investigate those subtle issues of the cosmological stimulated emission, we first outlines the details of setup in Sec.\,\ref{sec: setup}, and then derives the evolution equations governing both the number density and internal quantum phase of squeezed-state gravitons using the Heisenberg equation in Sec.\,\ref{sec: Heisenberg equation}. Similar methods have also been applied to the study of particle production \cite{Emond:2018ybc,Moroi:2020bkq} and neutrino kinematics \cite{Volpe:2013uxl,Volpe:2015rla,Vaananen:2013qja,Serreau:2014cfa,Kartavtsev:2015eva}.
The derivation follows an approach similar to the quantum-mechanical formulation of the Boltzmann equation, with the primary difference being the ansatz for quantum states. The sensitivity to IR gravitons is subtle in an expanding universe, as the physical interpretation of gravitons on superhorizon scales remains unclear. To address this, we begin with a careful analysis in a flat background where the issue is absent in Sec.\,\ref{sec_flat}. Then, we will show that the back reaction of radiation to the background spacetime is essential when discussing the sizable stimulated emission. As an alternative approach, we analyze cosmological stimulated emission in an expanding universe under a reasonable parameter space in Sec.\,\ref{sec: FLRW}, where the initial graviton wavelength is not significantly larger than the Hubble horizon size, ensuring that uncertainties from large gauge transformation do not arise. We employ the averaged Heisenberg equation~(distinct from the usual Boltzmann equation based on the kinematical description) and compute the evolution of graviton number, comparing the results obtained in the in-in formalism \cite{Ota:2024idm}. When the interaction coupling parameter is small, this new method accurately reproduces the results obtained from perturbation theory based on the in-in formalism. However, as the parameter increases, we observe significant deviations from the previous one-loop results, this indicates that our approach extends beyond conventional perturbative methods. In Sec.\,\ref{Sec: IR_Reg}, to address the secular IR issue, we propose a regularization method to eliminate this secular growth with a special focus on the large gauge symmetry. Finally, we conclude and outline possible directions for future work in Sec.\,\ref{sec:conclusion}.

\section{Basic setup}\label{sec: setup}
Let us consider an FLRW background spacetime and its traceless transverse perturbations:
\begin{align}
	ds^2 &= a(\tau)^2 \left(-d\tau^2 +\gamma_{ij}dx^idx^j \right),
	\\
	\gamma_{ij} &\equiv \delta_{ij} + \frac{2}{M_{\rm pl}}h_{ij} + \frac{2}{M^2_{\rm pl}} h_{ik}h^k{}_j+\cdots,
	\\
	h^i{}_{i} &= \partial_i h^i{}_{j} = 0.\label{eq:const}
\end{align}
Here, Latin indices refer to spatial coordinates and are raised and lowered using the background spatial metric $\delta^{ij}$ and $\delta_{ij}$. 
$M_{\rm pl}$ denotes the reduced Planck mass, and $a(\tau)$ is the isotropic scale factor. Expanding the Einstein-Hilbert action to the second order in $h^i{}_{j}$ yields the action for the free graviton:
\begin{align}
	S &= \int d\tau L[h^i{}_j,h'^i{}_j,\tau],
	\\
	 L[h^i{}_j,h'^i{}_j,\tau] &\equiv  \frac{1}{2}\int  d^3 x \, a(\tau)^2 \Big[(h'^i{}_j)^2  - (\partial_k h^i{}_j)^2 \Big].\label{defLagran}
\end{align}
We consider a thermal free scalar field $\chi$ that is minimally coupled to the graviton. 
Interactions between $\chi$ and $h_{ij}$ arise from the kinetic term:
\begin{align}
	-\frac{1}{2} \int d^4x \, \sqrt{-g} \, g^{\mu\nu} \partial_\mu \chi \, \partial_\nu \chi 
	&\supset   M_{\rm pl}^{-1} \int d^4 x \, a^2 \, h^{ij} \partial_i \chi \, \partial_j \chi.\label{defintL}
\end{align}
It has been found that the four-point interaction is eliminated by perturbing a tadpole diagram; hence, we truncate the interaction Hamiltonian at this order~\cite{Ota:2023iyh}.

\medskip
Define the conjugate momentum as
\begin{align}
	\pi^j{}_i \equiv \frac{\delta L[h^i{}_j,h'^i{}_j,\tau]}{\delta h^i{}_j} = a^2 h'^j{}_i\,.
\end{align}
The Fourier expansion of these tensor perturbations is given by 
\begin{align}
	h^i{}_j (\tau,\mathbf{x}) &= \sum_{s=\pm} \int \frac{d^3 k}{(2\pi)^3} e^{i\mathbf{k} \cdot \mathbf{x}} e^{i}{}_{j}(\mathbf{k},s) h^{(s)}_{\mathbf{k}}(\tau)\,,\label{hijxpand}
	\\
	\pi^i{}_j (\tau,\mathbf{x}) &= \sum_{s=\pm} \int \frac{d^3 k}{(2\pi)^3} e^{i\mathbf{k} \cdot \mathbf{x}} e^{i}{}_{j}(\mathbf{k},s) \pi^{(s)}_{\mathbf{k}}(\tau)\,,
\end{align}
where the polarization tensors satisfy
\begin{align}
	e^{i}{}_{j}(\mathbf{k},s) \left(e^{i}{}_{j}(\mathbf{k},s')\right)^* &= \delta_{ss'}\,,
	\\
	\left(e^{i}{}_{j}(\mathbf{k},s)\right)^* &= e^{i}{}_{j}(-\mathbf{k},s) = e^{i}{}_{j}(\mathbf{k},-s)\,.
\end{align}
Next, we impose the canonical commutation relation for the Fourier modes:
\begin{align}
	\left[ h^{(s)}_{\mathbf{k}},  \pi^{(s')}_{\mathbf{k}'}\right] = i\hbar \delta^{ss'} (2\pi)^3 \delta(\mathbf{k}+\mathbf{k'})\,.
\end{align}
Hereafter, we set $\hbar =1$.

\medskip
Equation~\eqref{defLagran} yields the free Hamiltonian 
\begin{align}
	H_0[h^i{}_j, \pi^j{}_i ,\tau]
	= \frac{1}{2} \sum_{s} \int \frac{d^3 k}{(2\pi)^3}  \left( \frac{ \pi^{(s)}_{\mathbf{k}}   \pi^{(s)}_{-\mathbf{k}} }{a^2}  + a^2 k^2  h^{(s)}_{\mathbf{k}}  h^{(s)}_{-\mathbf{k}} \right)\,.
\end{align}
Using the \textit{instantaneous} annihilation operator~\cite{Kanno:2021vwu}
\begin{align}
	d^{(s)}_{\mathbf{k}} \equiv a(\tau )\sqrt{\frac{k}{2}} h^{(s)}_{\mathbf{k}} + \frac{i}{a(\tau) \sqrt{2k}} \pi^{(s)}_{\mathbf{k}},\label{defd}
\end{align}
one can diagonalize the free Hamiltonian operator as
\begin{align}
	H_0[h^i{}_j, \pi^j{}_i ,\tau] &=  \sum_{s} \int \frac{d^3 k}{(2\pi)^3} \frac{k}{2}\left(  d^{(s)}_{\mathbf{k}} d^{(s)\dagger}_{\mathbf{k}} +  d^{(s)\dagger}_{\mathbf{k}} d^{(s)}_{\mathbf{k}}  \right).
\end{align}
From this, one can define the instantaneous number operator:
\begin{align}
	N^{(s)}_{\mathbf{k}}(\tau) &\equiv   d^{(s)\dagger}_{\mathbf{k}}(\tau) d^{(s)}_{\mathbf{k}}(\tau),\label{defN}
\end{align}
which allows us to express the Hamiltonian as
\begin{align}
	H_0[h^i{}_j, \pi^j{}_i ,\tau]
	=  \sum_{s=\pm} \int \frac{d^3 k}{(2\pi)^3} k\,N^{(s)}_{\mathbf{k}}(\tau) +\mathcal{C}\,.
\end{align}
where the last term $\mathcal{C}$ represents a constant energy shift. In addition to the number operator, we introduce 
\begin{align}
	L^{(s)}_{\mathbf{k}}(\tau) \equiv    d^{(s)}_{\mathbf{k}}(\tau) d^{(s)}_{-\mathbf{k}}(\tau)\,.
\end{align}
The expectation value of $L$ characterizes the coherence between different number eigenstates, which does not appear in standard kinetic theory formulated in terms of separable number eigenstates.

\section{Heisenberg equation}\label{sec: Heisenberg equation}
In this section, we derive the Heisenberg equation averaged over given quantum states. 
We do not call this equation Boltzmann equation to emphasize that we retain the quantum field theory perspective without fully reducing the system to kinetic theory.
The following equations describe the equations of motion for field operators in QFT where the ``size'' of each graviton is too large to regard them as point particles.

\subsection{Free theory}

Let us start with the free theory.
The free Heisenberg equation is given by
\begin{align}
	\frac{d  {\mathcal O}}{d\tau} = i [ H_0, {\mathcal O}] + \frac{\partial {\mathcal O}}{\partial \tau}\,.
\end{align}
In an expanding universe, operators depend explicitly on time via the scale factor, which gives rise to the last term.

Using Eq.~\eqref{defd}, we obtain 
\begin{align}
		\left[ H_{0}, N^{(s)}_{\mathbf{k}}\right] &= 0\,,
	\\
	\left[ H_{0}, L^{(s)}_{\mathbf{k}}\right] &= - 2k  L^{(s)}_{\mathbf{k}}\,.
\end{align}
Next, to evaluate the partial derivatives of the operators, we compute
\begin{align}
	\frac{\partial}{\partial \tau} d_{\mathbf{k}}^{(s)}(\tau) 
	&= \frac{1}{a(\tau)} \frac{\partial a(\tau )}{\partial \tau} \left[ a(\tau) \sqrt{\frac{k}{2}} h^{(s)}_{\mathbf{k}} - \frac{i}{a(\tau) \sqrt{2k}} \pi^{(s)}_{\mathbf{k}}\right]
			\notag 
	\\
	&= \frac{1}{a(\tau)} \frac{\partial a(\tau )}{\partial \tau} \left[ a(\tau) \sqrt{\frac{k}{2}} h^{(s)}_{-\mathbf{k}} + \frac{i}{a(\tau) \sqrt{2k}} \pi^{(s)}_{-\mathbf{k}}\right]^\dagger.
\end{align}
Note that the canonical variables do not explicitly depend on time.
Summarizing, with $\mathcal{H} \equiv \partial_\tau a/a$, we obtain
\begin{align}
	\frac{\partial}{\partial \tau} d_{\mathbf{k}}^{(s)}(\tau) &= \mathcal{H} d_{- \mathbf{k}}^{(s)\dagger}(\tau)\,,
	\\
	\frac{\partial}{\partial \tau} d_{\mathbf{k}}^{(s)\dagger}(\tau) &= \mathcal{H} d_{- \mathbf{k}}^{(s)}(\tau)\,.
\end{align}
Using these, we find
\begin{align}
\begin{split}
	\frac{d  N^{(s)}_{\mathbf{k}}}{d\tau} &=  \mathcal{H} \left(  L^{(s)}_{\mathbf{k}} +    L^{(s)\dagger}_{\mathbf{k}}\right)\,,
	\\
	\frac{d  L^{(s)}_{\mathbf{k}}}{d\tau} &= -2ik  L^{(s)}_{\mathbf{k}}+   \mathcal{H} \left(   N^{(s)}_{\mathbf{k}} +  N^{(s)}_{-\mathbf{k}} +V \right)\,,\label{freeeom}
\end{split}
\end{align}
\medskip
Now, consider the initial vacuum state $ \varrho_0 = |0\rangle \langle 0|$ and expand the instantaneous ladder operator with respect to this initial vacuum state: 
\begin{align}
	d^{(s)}_{\mathbf{k}} (\tau)= \mu_k(\tau,\tau_0)  d^{(s)}_{\mathbf{k}} (\tau_0)+ \nu_k (\tau,\tau_0) d^{(s)\dagger}_{ -\mathbf{k}}(\tau_0)\,.
\end{align}
Taking the expectation value of Eq.~\eqref{freeeom} with respect to $ \varrho_0$, we obtain
\begin{align}
	\frac{d |\nu_k|^2}{d\tau} &=  \mathcal{H} \left( \mu_k\nu_k  +   \mu_k^*\nu_k^*\right)\,,
	\\
	\frac{d (\mu_k\nu_k) }{d\tau} &= -2 i k  \mu_k \nu_k +   \mathcal{H} \left( 2|\nu_k^2| + 1 \right)\,.
\end{align}
These equations are consistent with the equation of motion for the mode functions, as the Bogoliubov coefficients depend on time through the canonical variables.

\subsection{Interaction theory}

In the interaction picture, the field operators remain free, while the density operator follows the von Neumann equation with respect to the interaction Hamiltonian $H_I$:
\begin{align}
	\frac{d  {\varrho}}{d\tau} = -i [ H_I, \varrho]\,. 
\end{align}
The evolution equation for the expectation value of ${\mathcal O}$ is  
\begin{align}
	\frac{d}{d\tau}\text{Tr}\left[ \varrho {\mathcal O}\right] &= \text{Tr}\left[ \frac{d \varrho }{d\tau} {\mathcal O} +  \varrho \frac{d{\mathcal O}}{d\tau}\right]
	\notag\\
	 &= \text{Tr}\left[-i[ H_I, \varrho] {\mathcal O} + i \varrho \left[  H_0, {\mathcal O}\right] +  \varrho \frac{\partial {\mathcal O}}{\partial \tau} \right] 
	 \notag\\
	 &=\text{Tr}\left[ i \varrho [ H_I, {\mathcal O}] + i \varrho [ H_0 , {\mathcal O}] +  \varrho \frac{\partial {\mathcal O}}{\partial \tau}  \right].\label{eq000}
\end{align}
The first equality follows from the linearity of the trace and the Leibniz rule for the total derivative.
Next, we define
\begin{align}
	n&\equiv \frac{1}{V}\text{Tr}\left[ \varrho  N_{\mathbf k}^{(s)}\right],
	\\
	\lambda &\equiv \frac{1}{V}\text{Tr}\left[ \varrho  L_{\mathbf k}^{(s)}\right] ,
\end{align}
where $V=(2\pi)^3\delta^{(3)}(0)$. In the absence of interactions, these reduce to $n_{0}  = |\nu_k|^2$ and  $\lambda_{0}  =\mu_k\nu_k $, with the subscript implying the order in the interaction Hamiltonians. 
Since $\varrho$ includes the effect of interactions to all orders, so do $n$ and $\lambda$. The total time derivatives are given by
\begin{align}
	\frac{1}{V}\frac{d}{d\tau}\text{Tr}\left[ \varrho  N_{\mathbf k}^{(s)}\right] &= \frac{dn}{d\tau}\,,
	\\
	\frac{1}{V}\frac{d}{d\tau}\text{Tr}\left[ \varrho  L_{\mathbf k}^{(s)}\right] &= \frac{d\lambda}{d\tau}\,. 
\end{align}
Moreover, we have
\begin{align}
\frac{1}{V}\text{Tr}\left[ \varrho	\,	i \left[ H_{0}, N^{(s)}_{\mathbf k}\right]\right]
	&=0\,,
	\\
\frac{1}{V}\text{Tr}\left[ \varrho	\,i \left[ H_{0}, L^{(s)}_{\mathbf k}\right]\right]
	&=- 2i k  \lambda\,,\label{eq3}
\end{align}
and also
\begin{align}
	\frac{1}{V}\text{Tr}\left[ \varrho	 \frac{\partial  N^{(s)}_{\mathbf k}}{\partial \tau} \right]&=  2 \mathcal H  \text{Re}[\lambda]\,,\label{eq1}
	\\
	\frac{1}{V}\text{Tr}\left[ \varrho	 \frac{\partial   L^{(s)}_{\mathbf k}}{\partial \tau} \right]&=    \mathcal H \left(   2 n + 1 \right )\,.\label{eq2}
\end{align}
Thus, the last two terms in Eq.~\eqref{eq000} are evaluated non-perturbatively.
In Ref.~\cite{Moroi:2020bkq}, particle production from a scalar condensate was considered for $ H_I = \bar \phi(t)\chi\chi$.
In that case, the interaction term was also evaluated non-perturbatively, which differs from the setup here.

\medskip
In general, a perturbative expansion is needed for the interaction term.
The interaction-state evolution is given by $\varrho \to F \varrho_0 F^{-1}$ with the initial state $\varrho_0$ and the evolution operator
\begin{align}
	F \equiv  T \exp \left(  -i \int^\tau_0d\tau' H_I(\tau')      \right)\,.
\end{align}
Then, we obtain
\begin{align}
	\text{Tr}\left[ i \varrho [ H_I, {\mathcal O}]\right] = \text{Tr}\left[ i \varrho_0 F^{-1} [ H_I, {\mathcal O}] F\right]\,,
\end{align}
which can be expanded as~\cite{Weinberg:2005vy}
\begin{align}
	\text{Tr}\left[ i \varrho [ H_I, {\mathcal O}]\right] = \sum_{n=0}^\infty i^n \int^\tau d\tau_1 \cdots \int^{\tau_{n-1}}d\tau_n  {\rm Tr}\left[\varrho_0 \left[ H_I(\tau_n) , \cdots \left[ H_I(\tau_1) ,i [ H_I, {\mathcal O}] \right]\right]\right]\,.\label{ininform2}
\end{align}
Under the Born-Markov approximation~\cite{Zarei:2021dpb}, the interaction term simplifies to
\begin{align}
	\text{Tr}\left[ i \varrho [ H_I, {\mathcal O}]\right] \simeq   {\rm Tr}\left[i \varrho_0 [ H_I, {\mathcal O}] \right] -  \int^\infty_0 d\tilde \tau   {\rm Tr}\left[\varrho_0 \left[ H_I(\tau - \tilde \tau) ,[ H_I(\tau), {\mathcal O}(\tau)] \right]\right]\,.\label{ininform4}
\end{align}

\subsection{Backward scattering}

The forward scattering vanishes for the 3-point interaction, so we only need to evaluate the second term in Eq.~\eqref{ininform4}.  
From Eq.~\eqref{defintL}, the interaction Hamiltonian is given by  
\begin{align}
		{H}_I &= -\frac{a^2}{M_{\rm pl}} \int d^3x \, {h}^{ij} \partial_i {\chi} \, \partial_j {\chi}\,. 
\end{align}
Using Eq.~\eqref{hijxpand} and  
\begin{align}
		\chi (\tau,\mathbf x)&= \int \frac{d^3 k}{(2\pi)^3} e^{i\mathbf k\cdot \mathbf x} \chi_{\mathbf k}(\tau)\,,
\end{align}
we can rewrite the interaction Hamiltonian as  
\begin{align}
	{H}_I 
	&= \frac{a^2}{M_{\rm pl}} \int \frac{d^3k \, d^3p_1 \, d^3p_2}{(2\pi)^9} (2\pi)^3 \delta(\mathbf{k} + \mathbf{p_1} + \mathbf{p_2}) \sum_{s} e^{(s)}_{ij}({\mathbf{k}}) {h}^{(s)}_{\mathbf{k}} p_{1i} p_{2j} {\chi}_{\mathbf{p_1}} {\chi}_{\mathbf{p_2}}\,.
\end{align}
Let us define the energy-momentum tensor projected onto the graviton polarization plane as  
\begin{align}
    {T}^{(s)}_{\mathbf{k}} &\equiv -a^2 \int \frac{d^3 l \, d^3 p}{(2\pi)^3} \delta(\mathbf{k} + \mathbf{l} + \mathbf{p}) e^{(s)}_{ij}({\mathbf{k}}) l_i p_j {\chi}_{\mathbf{l}} {\chi}_{\mathbf{p}}\,. \label{defemts}
\end{align}
Then, the interaction Hamiltonian can be further simplified to  
\begin{align}
    {H}_{I} = -M_{\mathrm{pl}}^{-1} \sum_{s=\pm} \int \frac{d^3 k}{(2\pi)^3} \, {h}^{(s)}_{\mathbf{k}} {T}^{(s)}_{\mathbf{k}}\,, \label{intHdef3}
\end{align}
or more concisely,  
\begin{align}
    {H}_{I} = -  M_{\mathrm{pl}}^{-1} {h}^S {T}_S. \label{intHdefsimp}
\end{align}
Here, $S$ encapsulates both the momentum and polarization indices and repeated $S$ implies summation over polarization and integration over momentum.  
We will use the notations in Eqs.~\eqref{intHdef3} and \eqref{intHdefsimp} interchangeably in the following discussion.  

For $\mathcal O =N^S, L^S$, Eq.~\eqref{ininform4} expands as 
\begin{align}
	 -  \int^\infty_0 d\tilde \tau    \left[ H_I(\tau - \tilde \tau) ,[ H_I(\tau), {\mathcal O}(\tau)] \right] =   A[\mathcal O] +   B[\mathcal O],
\end{align}
where  	
\begin{align}	
    A[\mathcal O] &=
    - \frac{1}{2  M^2_{\rm pl}} \int^{\infty}_0 d\tilde \tau \,   
     \left[ {T}_{S_2}(\tau-\tilde \tau), {T}_{S_1}(\tau) \right] 
       \nonumber\\
    &\qquad\qquad\qquad\times\Big\{{h}^{S_2}(\tau-\tilde \tau) \left[ {h}^{S_1}(\tau), \mathcal O(\tau) \right]+\left[ {h}^{S_1}(\tau), \mathcal O(\tau) \right] {h}^{S_2}(\tau-\tilde \tau)\Big\}\,, \label{Bogotransf2}
    \\	
    B[\mathcal O] &=
    - \frac{1}{2  M^2_{\rm pl}} \int^{\infty}_0 d\tilde \tau  \left[{h}^{S_2}(\tau-\tilde \tau), \left[ {h}^{S_1}(\tau), \mathcal O(\tau) \right]\right]\nonumber\\
     &\qquad\qquad\qquad\times\Big\{{{T}_{S_2}(\tau-\tilde \tau) {T}_{S_1}(\tau) + {T}_{S_1}(\tau) {T}_{S_2}(\tau-\tilde \tau)}\Big\}\,.\label{Bogotransf4}
\end{align}	
The term $B[\mathcal O]$ is independent of the graviton quantum state, corresponding to the spontaneous emission from the scalar field.  
This type of contribution is often considered in the graviton production in thermal plasma~\cite{Ghiglieri:2015nfa} and induced gravitational waves from density fluctuations~\cite{Saito:2008jc}.
In contrast, $A[\mathcal O]$ depends on the initial graviton state and thus represents stimulated emission.  
One can show that $A[\mathcal O]$ dominates for infrared (IR) graviton modes. Therefore, we will neglect spontaneous emission in the subsequent analysis.

\subsection{Stimulated emission}

The evaluation of Eq.~\eqref{Bogotransf2} is similar to that in Ref.~\cite{Ota:2024idm}, and it is simplified due to the Markov approximation.  
Using the retarded and Keldysh propagators,  
\begin{align}
G^R_{p_1}(\tau_1, \tau_2) (2\pi)^3 \delta(\mathbf{p_1} + \mathbf{p_2}) &= 	i a^2(\tau_2) \Theta(\tau_1 - \tau_2) [{\chi}_{\mathbf{p_1}}(\tau_1), {\chi}_{\mathbf{p_2}}(\tau_2)]\,, \label{defretG}
\\
G^K_{l_1}(\tau_1, \tau_2) (2\pi)^3 \delta(\mathbf{l_1} + \mathbf{l_2}) &=	a^2(\tau_2) \, \text{Tr}\left[{\varrho} ({\chi}_{\mathbf{l_2}}(\tau_2) {\chi}_{\mathbf{l_1}}(\tau_1) + {\chi}_{\mathbf{l_1}}(\tau_1) {\chi}_{\mathbf{l_2}}(\tau_2))\right]_\chi\,, \label{defKelG}
\end{align}
we find  
\begin{align}
	\text{Tr}\left[{\varrho}_\chi \left[ {T}_{S_2}(\tau-\tilde \tau), {T}_{S_1}(\tau) \right]\right] &= 2i\,\delta_{s_1s_2} (2\pi)^3 \delta(\mathbf{k_1} + \mathbf{k_2}) X(\tau,\tau-\tilde \tau)\,,\label{expTderive2}
\end{align}
where we define  
\begin{align}
	X(\tau_1,\tau_2) \equiv   \frac{2}{15} \frac{a^2(\tau_1)}{a^2(\tau_2)} \int \frac{p^2 dp}{2\pi^2} p^4 G^K_{p}(\tau_1, \tau_2) G^R_{p}(\tau_1, \tau_2)\,. \label{expTderive:defX}
\end{align}
Since the thermal scalar field is well inside the horizon, the propagators take the Minkowski form rescaled by the scale factor:  
\begin{align}
	G^R_{p}(\tau_1,\tau_2) &= \frac{a(\tau_2)}{a(\tau_1)} \frac{\sin p(\tau_1-\tau_2)}{p}  \Theta(\tau_1-\tau_2)\,,\label{defr}
	\\
	G^K_{p}(\tau_1,\tau_2) &\simeq  \frac{a(\tau_2)}{a(\tau_1)}\frac{\cos p(\tau_1-\tau_2)}{p} (1 + 2 f_{\beta p})\,, \label{defk}
\end{align}
where $f_{\beta p} = \frac{1}{e^{\beta p} - 1}$ stands for the thermal distribution of bosonic particles, with $\beta$ denoting the inverse of the comoving temperature. Thus, the scale factor cancels out, and $X$ becomes a function of $\tilde \tau$:  
\begin{align}
X(\tilde \tau) = \frac{1}{15 \pi^2} \left[- \frac{3}{8 \tilde \tau^5} + \text{csch}^5\left(\frac{2 \pi \tilde \tau}{\beta}\right)  \frac{\pi^5}{\beta^5} \left(11 \cosh\left(\frac{2 \pi \tilde \tau}{\beta}\right)
    + \cosh\left(\frac{6 \pi \tilde \tau}{\beta}\right)\right)  \right]\,.
\end{align}
$X(\tilde \tau)$ acts as a window function open for $\tilde \tau \sim \beta/(2\pi)$ and vanishes for $\tilde \tau/\beta \gg 1$, justifying the Markov approximation.  
By integrating $X(\tilde \tau)$, we obtain  
\begin{align}
    \lim_{\beta / \tau \to 0} \int^{\infty}_0 d\tilde \tau \, X(\tilde \tau) = \frac{\pi^2}{450 \beta^4}\,.
\end{align}
Thanks to the Markov approximation, the graviton operators reduce to a local operator product:  
\begin{align}
	  A[\mathcal O] &\simeq 
    - \frac{i \pi^2}{450 M_{\rm pl}^{2}\beta^4} \delta_{S_1S_2} \Big\{{h}^{S_2}(\tau) \left[ {h}^{S_1}(\tau), \mathcal O(\tau) \right]+\left[ {h}^{S_1}(\tau), \mathcal O(\tau) \right] {h}^{S_2}(\tau) \Big\}\,. \label{Bogotransf22}
\end{align}
We have also verified that the non Markovian correction is $\mathcal O(\beta^{-3})$, which we may ignore in the present case.
Using the instantaneous operator~\eqref{defd}, the field operator takes the form  
\begin{align}
	h^{(s)}_{\mathbf k} = \frac{1}{a}\frac{1}{\sqrt{2k}}\left(  d^{(s)}_{\mathbf k}  +   d^{(s)\dagger}_{-\mathbf k}\right)\,,
\end{align}
which yields  
\begin{align}
\big\langle 0\big|	{h}^{S_2} \left[ {h}^{S_1}, N^S \right]+\left[ {h}^{S_1}, N^S \right] {h}^{S_2} \big| 0 \big\rangle
	& =\frac{V}{ka^2}\delta^{SS_1}\delta^{SS_2}\left(   \lambda^{S_2}_0 - \lambda^{S_2*}_0   \right)\,,
	\\
	\big\langle 0\big|	{h}^{S_2} \left[ {h}^{S_1}, L^S \right]+\left[ {h}^{S_1}, L^S \right] {h}^{S_2} \big| 0 \big\rangle
	& =-\frac{V}{ka^2}\delta^{SS_1}\delta^{SS_2}\left(   2 \lambda^{S_2}_0 + 2 n^{S_2}_0 +1  \right)\,,
\end{align}
where the repeated $S$ indices on the LHS are not summed.  
Contracting $S_1$ and $S_2$, we obtain  
\begin{align}
	V^{-1}  A[N^S] &= \frac{ \pi^2}{225 M_{\rm pl}^{2}ka^2\beta^4} \text{Im} [ \lambda^{S}_0]\,,\label{eq4}
	\\
	V^{-1}  A[L^S] &= i \frac{ \pi^2}{225 M_{\rm pl}^{2}ka^2\beta^4} \left( \lambda^{S}_0 + n^{S}_0 + \frac{1}{2}  \right)\,. \label{eq5}
\end{align}
Due to the Markov approximation, terms of $\mathcal{O}(1/(M_{\rm pl}\beta))$ are neglected, which is justified unless the system is near the Planck scale.

\subsection{Summary so far}\label{subsecsum}

Combining Eqs.~\eqref{eq3}-\eqref{eq2} and \eqref{eq4}-\eqref{eq5}, as well as elevating free variables to their non-perturbative counterparts by matching their initial values, we obtain the Heisenberg equations averaged over a general quantum state: 
\begin{align}
	n' &= 2\frac{a'}{a}  \text{Re}[\lambda] + \alpha \text{Im} [ \lambda]\,, \\
	\lambda' &=- 2i \lambda  +  2\frac{a'}{a} \left(   n + \frac{1}{2} \right ) +i \alpha \left( \lambda + n + \frac{1}{2}  \right)\,,
\end{align}
where we define the dimensionless time variables normalized by $k$:  
$x = k\tau$, $x_\beta = k\beta$, $x_{\rm pl} = k/M_{\rm pl}$, and $'\equiv d/dx$, along with  
\begin{align}
	\alpha \equiv \frac{\pi^2 x_{\rm{pl}}^2}{225 a^2x^4_\beta}\,.
\end{align}
Note that in the interaction terms in Eqs.~\eqref{eq4} and \eqref{eq5}, $n$ and $\lambda$ correspond to their free values indexed as 0, while they can be identified with their non-perturbative counterparts at the initial time.
At each time step, the perturbed values of $n$ and $\lambda$ are computed and used as initial conditions for the next step.
This iterative procedure continues until the final values are determined.
A similar prescription is employed in deriving the standard Boltzmann equations based on the Born-Markov approximation.

To simplify the equations further, we separate the real and imaginary parts of variable $\lambda$ as 
\begin{align}
	\lambda = \xi + i \eta.
\end{align}
where $\xi$ and $\eta$ are 
real numbers. This leads to the following set of equations:
\begin{keyeqn}
\begin{align}
	n' &= 2\frac{a'}{a}  \xi + \alpha \eta\,, \label{rh1}\\
	\xi' &= 2 \eta  +  2\frac{a'}{a} \left(   n + \frac{1}{2} \right ) -  \alpha \eta\,, \label{rh2}\\
	\eta' &=- 2\xi    +  \alpha \left( \xi + n + \frac{1}{2}  \right)\,. \label{rh3}
\end{align}
\end{keyeqn}
For example, during the radiation dominant era, we have $a'/a=1/x$.

\section{Flat Background} \label{sec_flat}

\begin{figure}
	\centering
        \hspace{-0.5cm}
	\includegraphics[width=\linewidth]{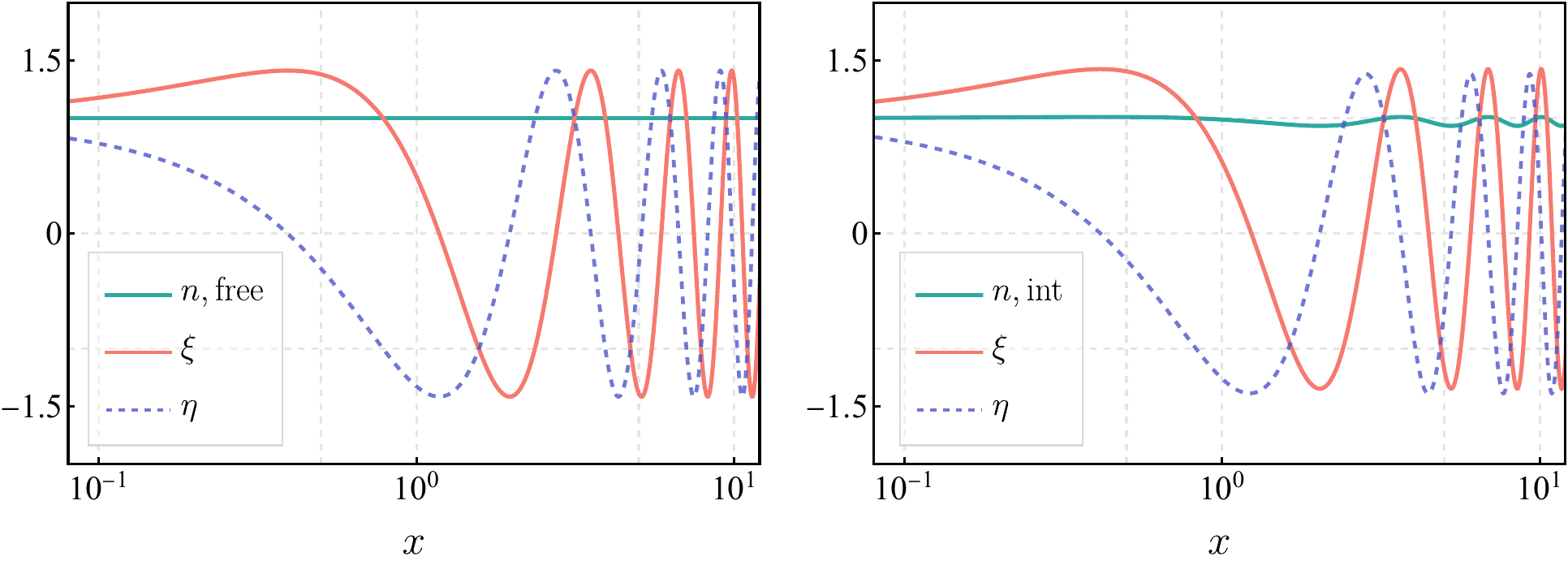}
	\caption{\textit{Left}: Free evolution of $n$~(\textcolor[RGB]{46,169,157}{green}), $\xi$~(\textcolor[RGB]{246,122,111}{red}), $\eta$~(\textcolor[RGB]{122,119,213}{blue dashed}).
	\textit{Right:} Interacting evolution of of $n$~(\textcolor[RGB]{46,169,157}{green}), $\xi$~(\textcolor[RGB]{246,122,111}{red}), $\eta$~(\textcolor[RGB]{122,119,213}{blue dashed}), with the parameters $x_{\rm pl}=10^{-41}$, $x_{\beta}=3\times10^{-21}$ and the initial conditions are set to $n=1$, $\theta=\pi/4$.}
	\label{fig1}
\end{figure}

While the above formulation applies to a general FLRW background, we first consider a simple Minkowski background to examine the physical properties of graviton emission.
In the Minkowski limit, we set $a=1$.
Here, $\beta^{-1}$ represents the physical temperature, and $\tau$ becomes 
the physical time.
The emission rate then simplifies to
\begin{align}
	\alpha  = \frac{\pi^2 x_{\text{pl}}^2}{225 x_\beta^4}\,.
\end{align}
Since the Hubble parameter vanishes in this limit, the Heisenberg equations~\eqref{rh1}--\eqref{rh3} reduce to 
\begin{align}
		\frac{d n}{dx} &=  \alpha \eta\,, \label{1steqmink} \\
		\frac{d \xi}{dx} &= 2  \eta - \alpha \eta\,, \\
		\frac{d \eta}{dx} &=  -2  \xi + \alpha \left( n + \xi+ \frac{1}{2} \right)\,.\label{3rdeqmink}
\end{align}
A general squeezed vacuum state initial condition is given by
\begin{align}
	\xi(0) &=\sqrt{n(0)(n(0)+1)}\cos\theta(0)\,, \label{uni1}\\
	\eta(0) &=\sqrt{n(0)(n(0)+1)}\sin\theta(0)\,\label{uni2}.
\end{align}
We recall that $\xi$ and $\eta$ represent the real and imaginary parts of $\lambda \sim \text{Tr}\left[\varrho L_{\mathbf{k}} \right]$, satisfy the relation:
$|\lambda|^2 = n(n+1)$,
where $n$ is the average particle number. Although we impose the above initial condition as an illustrative example, the effects discussed here are not restricted to the squeezed vacuum state. In fact, any initial conditions corresponding to more general initial states can be considered. In a Minkowski background, the vacuum state is uniquely defined as $n(0)=0$. 
For $n(0)\geq 1$ and arbitrary $\theta$, the states correspond to excited states, which can be expressed as squeezed vacuum states.
Without interaction with the scalar field ($\alpha=0$), $n_0$ remains constant, and $\lambda_0$ evolves as $\lambda_0 = e^{-2ix} \lambda_0(0)$.
Thus, the graviton number remains independent of the internal phase ($i.e.$ the intrinsic phase of $\lambda$).
In the left panel of Fig.~\ref{fig1}, we plot the time evolution of $n_0$, $\xi_0$, and $\eta_0$ for $n_0(0)=\xi_0(0)=\eta_0(0)=1$. 

When turning on interactions ($\alpha\neq0$), the internal phase and number are coupled.
For a typical parameter choice relevant to LIGO/Virgo-like gravitational wave interferometers~\cite{LIGOScientific:2016aoc}, we take:
\begin{itemize}
	\item $\ell\sim k^{-1}=\mathcal{O}(10^6)m$: the LIGO-Virgo detectable scale is $\mathcal O(100)$Hz.
	\item $x_{\rm pl} = k/M_{\rm pl}= 10^{-41}$. 
	\item $x_\beta = \frac{k}{M_{\rm pl}} \beta M_{\rm pl}\sim 10^{-41}\times10^{20}=10^{-21}$: $\beta^{-1} = \mathcal{O}(10^{-1})$ GeV.
\end{itemize}
The numerical result, shown in the right panel of Fig.~\ref{fig1}, demonstrates a small fluctuation in $n$.
Oscillations in the phase transfer into variations in the graviton number via Eq.~\eqref{1steqmink}.
In the right panel of Fig.~\ref{fig1}, the periodic average value of $n$ decreases, corresponding to stimulated absorption.
Depending on the initial phase $\theta$, the effect can manifest as either emission or absorption.

In the flat background case, the system of first-order differential equations (\ref{1steqmink})-(\ref{3rdeqmink}) can be solved exactly. Given the initial conditions $n(0)=1$ and $\theta=\pi/4$, the exact solution takes the form:
\begin{align}
    n(x)=\frac{\alpha}{2\sqrt{1-\alpha}}\sin\left(2 x\sqrt{1-\alpha}\right)+\frac{\alpha(5\alpha-4)}{8(\alpha-1)}\cos\left(2x\sqrt{1-\alpha}\right)+\frac{1}{8}\left(7-5\alpha+\frac{1}{1-\alpha}\right)\,.\label{solnmink}
\end{align}
When $\alpha<1$, the particle number oscillates over time and the behavior remains perturbative.
However, for $\alpha>1$, the trigonometric function reduce to their hyperbolic counterparts and then instability arises, causing the number density to grow exponentially. Neglecting the exponentially suppressed terms, one can approximately write $n(x)$ as
\begin{align}
    n(x)=\frac{1}{16}\left(5+\frac{4}{(\alpha-1)^{1/2}}+\frac{1}{\alpha-1}\right)e^{2\sqrt{\alpha-1}\,x}\,.
\end{align}
For small $\alpha$, our prediction aligns with the perturbation theory based on the in-in formalism.
Indeed, by expanding Eq.~(\ref{solnmink}) to the leading order in $\alpha$, it reproduces the analytical expression derived in previous work~\cite{Ota:2024idm} using the in-in formalism:  
\begin{align}
    n^{\text{PT,1-loop}} = - \alpha \sqrt{n_0(n_0+1)} \sin (x) \sin (x-\theta_0)\,, \label{rmin}
\end{align}
which indicates that our new method provides a resummation of the in-in formalism, see also Appendix~\ref{appA} for a summary of previous results.  

In Fig.~\ref{fig1}, we present the solutions for $\alpha \sim 0.054$, corresponding to $x_\beta = 3 \times 10^{-21}$.  
The limitation of the in-in formalism arises because the unperturbed operators in the integral evolve with $\alpha$.  
If their nonlinear evolution deviates significantly from their background values, the traditional perturbative estimation fails.  

\begin{figure}
	\centering
	\includegraphics[width=0.9\linewidth]{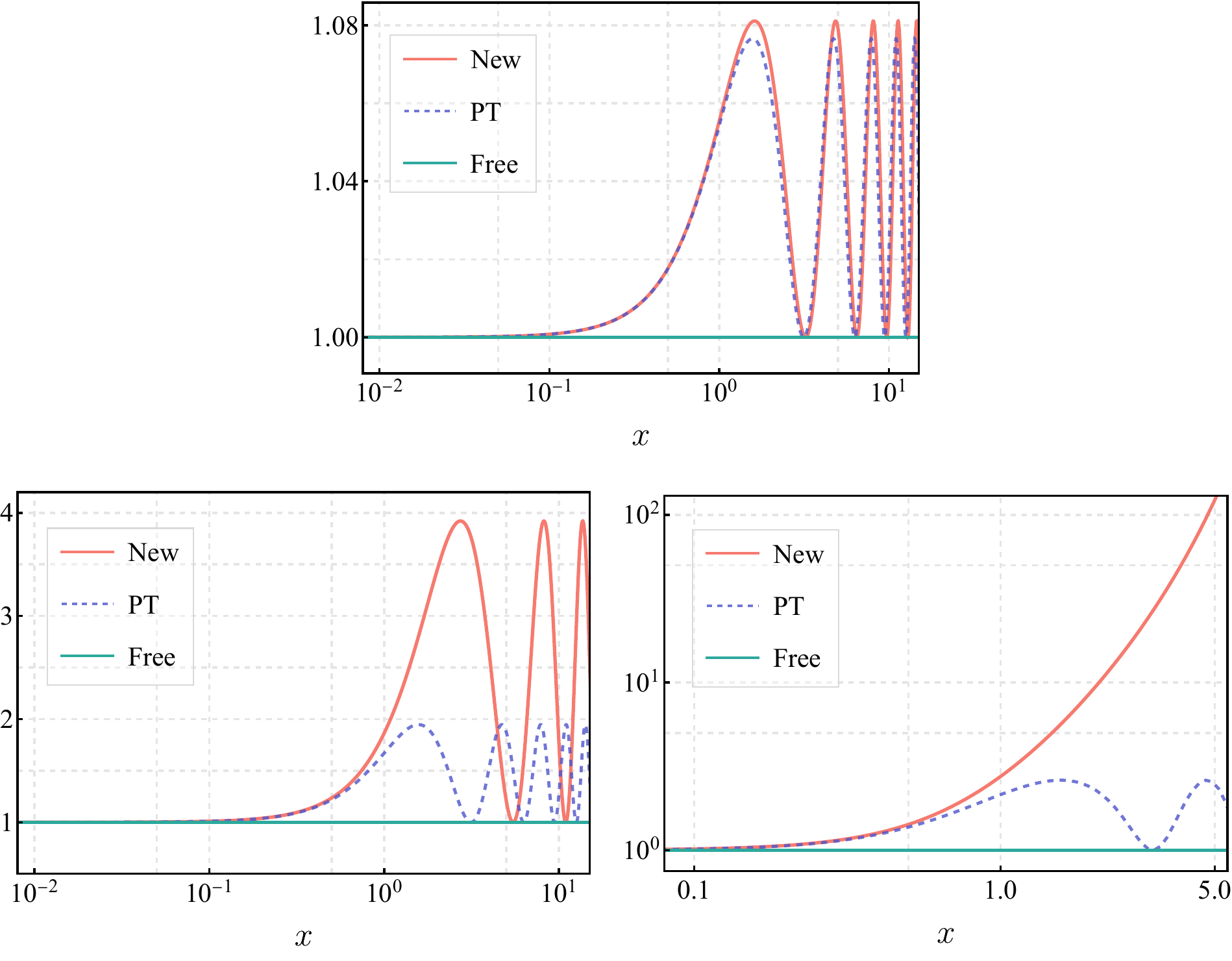}
	\caption{
    	The graviton number is computed using the Heisenberg equation~(\textcolor[RGB]{246,122,111}{red solid line}) and 1-loop perturbation theory~(\textcolor[RGB]{112,119,213}{blue dashed line}) in the Minkowski background. The initial conditions are set to $n(0)=1$, $\theta(0)=\pi$, and the parameter $x_{\rm pl} $ is set to $10^{-41}$.
	\textit{Top}: $x_{\beta}=3\times10^{-21}$, $i.e.$, $\alpha\sim 0.054$. The emission effect is small and remains perturbative.
	\textit{Bottom left}: $x_{\beta}=1.6\times 10^{-21}$, $i.e.$, $\alpha \sim 0.67$. The deviation between the Heisenberg equation and the 1-loop perturbation theory becomes apparent.
	\textit{Bottom right}: $x_{\beta}=1.4\times 10^{-21}$, $i.e.$, $\alpha \sim 1.14$, where a resonance effect is observed.  
	}
	\label{fig2}
\end{figure}

A comparison between 1-loop perturbation theory and the Heisenberg equation is shown in Fig.~\ref{fig2} for different choices of the parameter $\alpha$.  
When $\alpha$ is smaller than $\mathcal{O}(10^{-2})$, our method matches the previous one-loop perturbative calculation within the in-in formalism very well, with only a few percent difference.  
As $\alpha$ increases to $\mathcal{O}(10^{-1})$, the deviation becomes more significant, differing by an order of magnitude, suggesting that our approach extends beyond conventional perturbative methods.  
However, when $\alpha$ exceeds unity, the differential equations become unstable, indicating the breakdown of the Born approximation.  
Although we numerically compared our results with the in-in calculation up to one-loop order, the deeper connection between our method and the in-in formalism deserves further investigation, which we leave as a direction for future work.  

Higher temperatures or longer graviton wavelengths enhance stimulated effects.
As the temperature increases, the deviation from perturbation theory becomes more pronounced.
As we discussed before (\ref{solnmink}), the  stable regime corresponds to $\alpha<1$. 
For $\alpha>1$, the graviton number undergoes resonant exponential amplification. The resonance time scale is given by $(k\alpha)^{-1}$.

So far, we have considered a Minkowski background, treating the thermal scalar field as a spectator field.
In this setup, the thermal bath temperature is a free parameter.
However, the backreaction to spacetime dynamics cannot be neglected if the energy density $\rho_\chi = \pi^2/(30\beta^{4})$ is large.
Consider a thermal bath of size $\ell  = 2\pi \gamma/k$ with some numerical factor $\gamma$, and $k$ is the wavenumber of the emitted graviton.
For $\gamma>1$, a graviton can be enclosed within the bath.
The time scale $\ell$ is related to the curvature radius, and the physical Hubble parameter is given by $H=1/(2\ell)$.
To ensure that the backreaction remains small, we require 
\begin{align}
	3M_{\rm pl}^2 H^2 \gg \rho_\chi\,.
\end{align}
This condition can be rewritten as
\begin{align}
	\alpha \ll \frac{1}{40\pi^2 \gamma^2}\,.
\end{align}
Thus, $\alpha$ must remain at most at the sub-percent level to keep the backreaction to spacetime negligible.
Since $\alpha$ is determined by the graviton wavenumber and the thermal bath temperature, the stimulated emission effect remains well within the perturbative regime as long as the backreaction is insignificant.

Before closing this section, we offer several comments on the equations of motion we obtained.
Firstly, it turns out that $\lambda=0$ is a valid solution to the differential equations in the flat background, that corresponding to the expectation value of $L_{\mathbf{k}}$ in any particle number eigenstates.
Thus, one can safely assume that graviton mixed states can be expanded in terms of the Hamiltonian eigenstates.
With this initial condition, stimulated emission does not occur. 
In other words, assuming the graviton is initially in a mixed state that can be expanded in terms of the particle number eigenstates, stimulated emission will never occur, and $n$ will remain 
constant while $\lambda$ stays at zero.
Secondly, at next-to-leading order in the Markov approximation, a higher-order correction in $x_{\rm pl}/x_\beta$ appears as a time-ordered modification to Eq.~\eqref{Bogotransf22}. 
The correction to Eq.~\eqref{1steqmink} at this order corresponds to the $\Gamma_{\chi\chi h}$ term discussed in the introduction, with an overall minus factor.
This contribution generally leads to absorption, which persists even in mixed states.
However, since $x_{\rm pl}/x_\beta$ is typically very small, this effect remains negligible in most cases.

\section{FLRW Background}\label{sec: FLRW}

In the flat background, a higher thermal bath temperature or a longer graviton wavelength enhances stimulated graviton emission.
Such behavior is analogous to stimulated emission in quantum electrodynamics (QED).
In QED, the population of excited atomic electrons also decreases more significantly for lower atomic energy levels.
Thus, $\alpha$ has a cutoff, and the softest photon is constrained by the minimum energy gap in the atomic state that stimulates emission.

However, in the present case, $k$ and $\beta$ remain unconstrained.
In the previous section, we found that the temperature or wavelength must be sufficiently small to neglect the backreaction on spacetime dynamics.
At this stage, an FLRW background is essential for discussing significant stimulated emission.
As $\alpha$ increases, the backreaction of the thermal bath on spacetime dynamics becomes inevitable.
While such a situation is already interesting as a thought experiment, it is also of practical importance for gravitons predicted by inflation, as the scale-invariant spectrum is generated at the onset of the radiation-dominated universe.

In a general FLRW background, the quantities $\beta$, $\tau$, and $k$ are comoving.
If the thermal bath of $\chi$ drives the background evolution, the energy density of $\chi$ is given by
\begin{align}
	\rho_\chi = \frac{\pi^2}{30 (a\beta)^4}\,.
\end{align}
During radiation domination, the physical Hubble parameter $H$ and the scale factor $a$ satisfy the relation $aH = \tau^{-1}$.
Substituting this into the Friedmann equation
\begin{align}
	3M_{\rm pl}^2 H^2 = \rho_\chi\,,
\end{align}
we obtain the simple expression for the coefficient
\begin{align}
	\alpha \equiv \frac{2}{5x^2}\,,
\end{align}
where $x = k\tau$. The Heisenberg equations are then given by
\begin{align}
n' &=\frac{2\xi}{x}+ \frac{2\epsilon \eta}{5x^2}\,, 
\\
\xi' &=  2\eta + \frac{2}{x}\left(n+\frac{1}{2}\right)  - \frac{2\epsilon \eta}{5x^2}\,, 
\\
\eta' &= -2\xi + \frac{2\epsilon}{5x^2}\left( \xi + n +  \frac{1}{2}  \right)\,.
\end{align}
where we have introduced an order-counting parameter $\epsilon$, which denotes the strength of the interactions.
With $\epsilon=1$, stimulated emission dominates over the the Hubble friction term on superhorizon scales ($x\ll 1$).
This behavior has been reported in perturbative analyses based on in-in formalism \cite{Ota:2024idm}, where it results in secular, scale-invariant growth in the graviton power spectrum. This secular growth behavior has also been confirmed by recent work \cite{Frob:2025sfq}.
The breakdown of perturbative analysis was suggested on superhorizon scales.
Here, the Heisenberg equations incorporate nonlinear corrections in $\alpha$ as an accumulated effect, potentially leading to a partial resummation of the series in $\alpha$ as explicitly shown in the flat spacetime example in Sec.~\ref{sec_flat}.
 
\begin{figure}
	\includegraphics[width=1\linewidth]{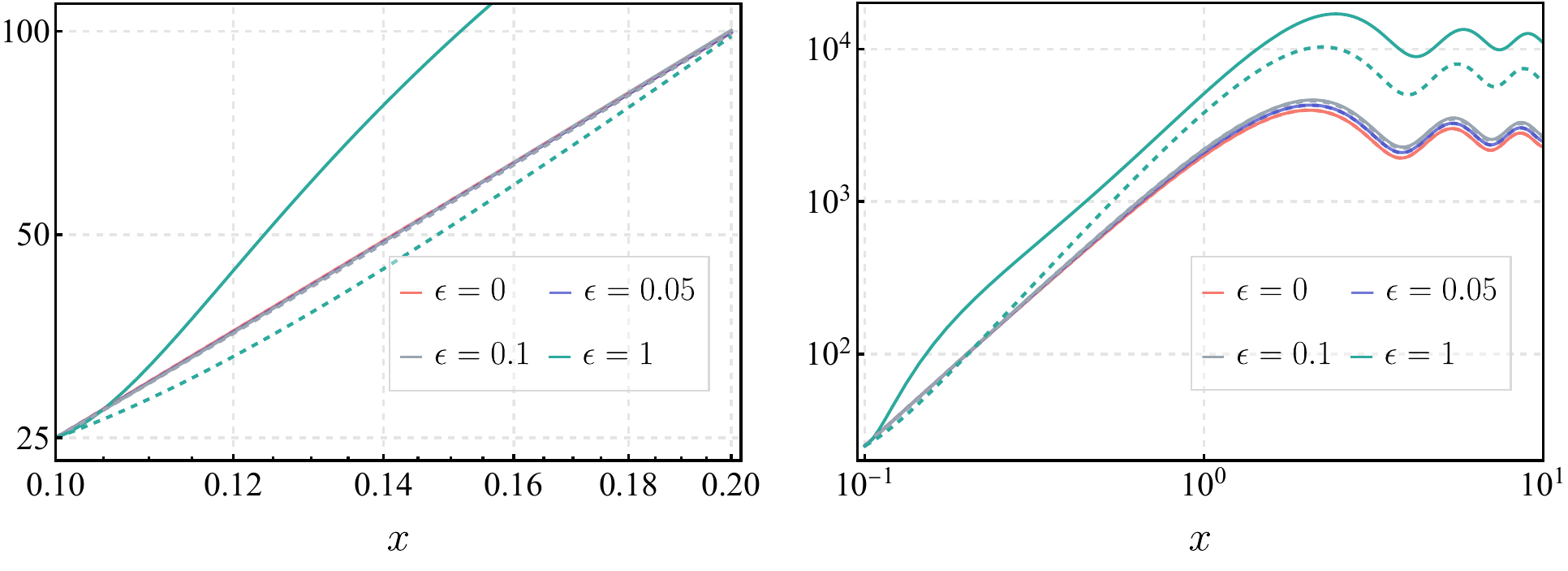}
	\caption{
	\textit{Left and Right}: $n$ as a function of $x=k\tau$ with $x_R=0.1$ for new (solid) and perturbation theory (dashed) solutions for $\epsilon=0, 0.05, 0.1, 1$. The left figure focuses on $x<0.2$.
	}
	\label{dig3}
\end{figure}

Assuming instantaneous reheating, if the interaction is turned off $\epsilon=0$, the expressions denote the free evolution for $n$ and $\lambda$, which are then given by~\cite{Ota:2024idm}
\begin{align}
  n_0(x) &= \frac{1}{8 x^2 x_R^4}\Bigg[1+2 x^2  + 2 x_R^4 +  
  2\left( x_R-x+2 x_R^2 x \right) \sin \big(2x - 2x_R\big)  \notag 
  \\
  &
  \qquad\quad\quad~~~+\big( 2 x_R ( x_R - 2 x ) - 1 \big) \cos\big( 2 x - 2 x_R\big) \Bigg]\,, 
  \label{nunu}\\
  \lambda_0(x) &= \frac{1}{8  x^2 x_R^4}\Bigg[ 
  + \Big( 4 x^2 x_R^2 + 4 i x_R\,x (x - x_R) 
  - 2 (x - x_R)^2 + 2 i x + 1 \Big) \cos\big( 2 x - 2 x_R\big) \nonumber\\
  &\quad+ 2 \left( x\left( x - i \right) 
  ( 2 x_R   - 2 i x^2_R  + i )  - x_R\right) \sin\big( 2x - 2x_R \big)-(2 i x + 1) \left( 2 x_R^4 + 1 \right) \Bigg]\,,
  \label{numu}
\end{align}
here $\tau_R$ denotes the reheating time, while $x_R=k\tau_R$ is the corresponding dimensionless variable.
These equations are derived by matching the inflationary mode functions with those during the radiation-dominated era. Consequently, the initial conditions for  the differential equations above can be chosen by evaluating the given expressions at $x_R$, yielding
\begin{align}
    n(x_R)=\frac{1}{4x^2_R}\,,\qquad\qquad \lambda (x_R)=\frac{1}{4 x^2_R}-\frac{i}{2 x_R}\,.
\end{align}
We illustrate the results for various values of $\epsilon$ in Fig.~\ref{dig3}.
For smaller $\epsilon$, the numerical solutions of the Heisenberg equations remain consistent with perturbation theory based on the in-in formalism.
On subhorizon scales, the dynamics also agree with perturbation theory except for their amplitudes, as the interaction terms vanish for large $x$.
On the super horizon scale, the interaction term becomes dominant, and exponential growth occurs, just like the case in the Minkowski background. 

As pointed out in Ref.~\cite{Ota:2023iyh}, the secular growth in the superhorizon limit is subtle.
Unlike the Minkowski background, superhorizon gravitons exhibit subtle behaviors.
In the linear theory, superhorizon tensor modes are constant, and they can be eliminated by a large gauge transformation, which is an asymptotic symmetry in the FLRW background~\cite{Weinberg:2003sw}.
Recently, it has been suggested that soft gravitons reside in a symmetric vacuum state, expressed as a superposition of infinitely degenerate vacua~\cite{Sloth:2025nan}. 
Eliminating soft gravitons implies that we take the symmetric vacuum with $n=0$ at the initial time; however, this is not a simple solution. 
After horizon re-entry, the graviton spectrum is expected to persist as a remnant of the primordial gravitons, requiring a return to the original symmetry-broken phase. 
Thus, the final spectrum depends on the choice of vacuum, and one must project the symmetric state onto the symmetry-broken vacuum. 
The challenge lies in the implementation of this projection, as stimulated emission becomes significant prior to horizon re-entry. 
Even if we choose $ n_0 = 0 $, corresponding to the symmetric vacuum, vacuum-stimulated emission still arises due to the vacuum fluctuations. This effect is then resonantly amplified, ultimately leading to secular growth in the IR limit.

The second issue concerns the robustness of perturbation theory.
One might think that the Heisenberg equation is useful because it can partially resum the loop spectrum at each time step, making it valid in the nonlinear regime, similar to the Boltzmann equation.
However, we argue that the Born approximation, assumed at the outset, may break down.
The interaction term is proportional to $x^{-2}$, suggesting a breakdown of perturbative expansion on superhorizon scales.
Rigorously speaking, we cannot justify the Born approximation in Eq.~\eqref{ininform4} without considering the other dimensional constants, while we can only do this after obtaining the Heisenberg equation.
This ambiguity arises from the dimensional coupling in general relativity.
Naively, the correction terms can be expanded in a series of $1/(\beta M_{\rm pl}) \sim \sqrt{H/M_{\rm pl}}$.
However, any negative power of $x$ may arise, which completely breaks perturbativity for $x<1$ at the level of the Heisenberg equation before the Born approximation.

\begin{figure}
	\includegraphics[width=\linewidth]{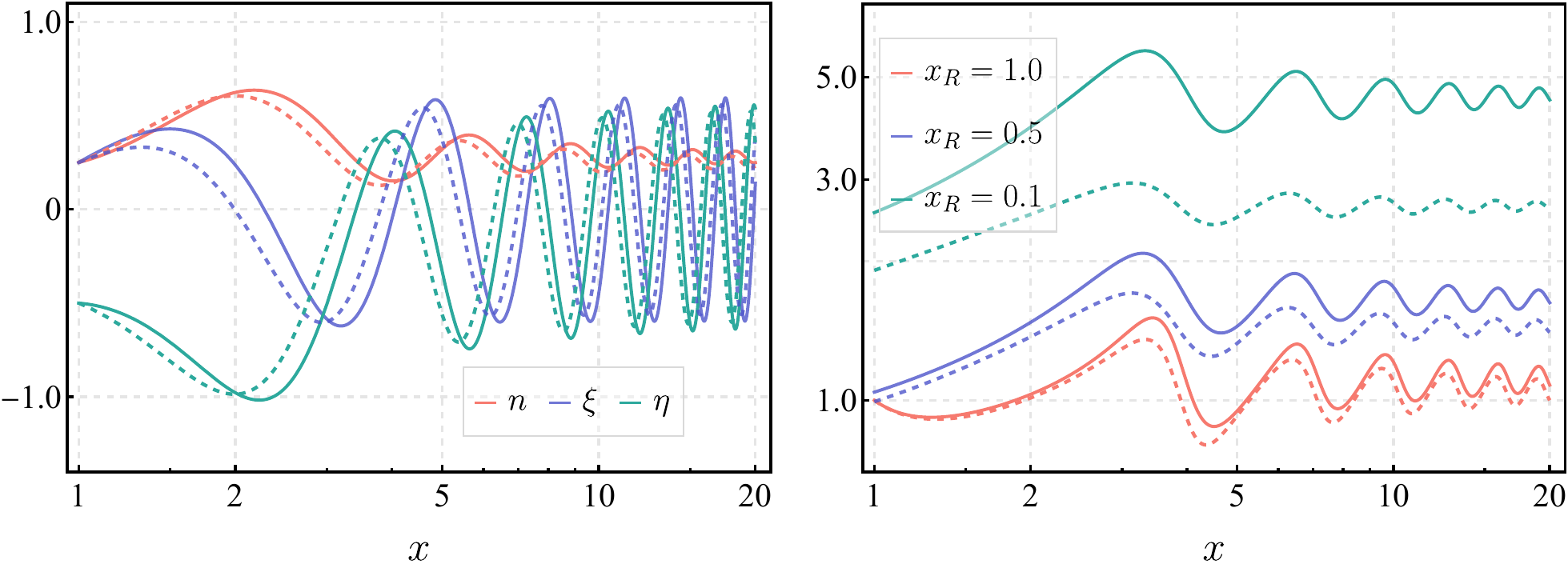}
	\caption{
	\textit{Left}: The evolution of $n$~(\textcolor[RGB]{247,122,111}{red}), $\xi$~(\textcolor[RGB]{77,85,204}{blue}), $\eta$~(\textcolor[RGB]{45,170,158}{green}) for free theory~(dashed) and interaction theory~(solid) with $\epsilon = 1$ and $x_R = k \tau_R = 1$.
	\textit{Right}: $n/n_0$ as a function of $x = k \tau$ with $\epsilon=1$, and $x_R = 1, 0.5, 0.1$ respectively. Solid lines represent the Heisenberg equation, and dashed lines represent perturbation theory results.
	}
	\label{digRR}
\end{figure}

As discussed above, the subtlety of gravitons in general relativity and the issue of perturbativity suggests that the reliable regime is only a few Hubble times after reheating, i.e., $x \gtrsim 1$.
In Fig.~\ref{digRR}, we illustrate the stimulated emission for $x_R=1$, which corresponds to the gravitons inside the horizon when inflation ends with instantaneous reheating.
In this regime, the issues discussed above are absent, and our equation of motion is reliable.
Even for $x_R = k \tau_R = 1$, stimulated emission is observed.
In the right panel of Fig.~\ref{digRR}, $n/n_0$ is shown as a function of $x=k\tau$ and $x_R = 1, 0.5, 0.1$. Solid lines represent the Heisenberg equation, and dashed lines represent perturbation theory.
As $x_R$ decreases, the initial gravitons become softer, leading to a larger stimulated emission.

Finally, we close this section with a comment.
Even without interaction ($\epsilon = 0$), $n$ and $\lambda = \xi + i \eta$ couple due to cosmic expansion, meaning that $\lambda = 0$ cannot be taken as a consistent ansatz in the expanding background where $x\sim 1$ is allowed. 
This indicates the failure of the assumption behind the standard kinetic theory, where $x\to \infty$ is implicit.
Quantum mechanical states are inevitably squeezed in a time-dependent background, even if they are initially separable.

\section{IR regularization and large gauge  transformation}\label{Sec: IR_Reg}

The secular growth in the graviton number indicates the superhorizon evolution of the IR mode~\cite{Ota:2023iyh}.
In this section, we discuss the IR dynamics and present a prescription for eliminating the secular term from a symmetry perspective.

Free IR graviton modes obey the linearized Einstein equation:
\begin{align}
	h_{ij}'' + 2 \frac{a'}{a}h_{ij}' =0.\label{eomIR}
\end{align}
This equation of motion is invariant under the large gauge transformation 
\begin{align}
	\xi^i = \frac{1}{2}\epsilon^{ij}x_j,
\end{align}
which turns into a constant shift in $h_{ij}$:
\begin{align}
	h_{ij} \to h_{ij} + \epsilon_{ij}.\label{cshift}
\end{align}
More precisely, $\epsilon_{ij}$ can correspond to either a constant or a decaying adiabatic mode~\cite{Pajer:2017hmb}.
The latter results in a constant shift for the conjugate momentum $\pi_{ij}$.
Since the constant mode can be eliminated by a large gauge transformation, free IR graviton modes are considered gauge-dependent.
As the secular term is due to the IR graviton modes, we need to understand if the effect is physical or not in our setup.

In practice, a gradient term emerges in Eq.~\eqref{eomIR} for subhorizon modes, explicitly breaking the large gauge symmetry.
Primordial gravitons are quantum-mechanically generated during inflation.
When generated, they were within the horizon and the adiabatic vacuum is chosen as a vacuum state in the remote past.
After the horizon exit, the quantum state remains adiabatic, which is regarded as a symmetry-broken phase in the IR regime.
This implies that the constant shift \eqref{cshift}, i.e., the residual gauge symmetry is regarded as a global symmetry, which can be spontaneously broken by the quantum state chosen by inflation.

In the present case, the stimulated emission is regarded as a tachyonic instability due to the negative mass squared.
In fact, one may consider an effective Hamiltonian 
\begin{align}
	H_{\rm eff} = -\frac{1}{5}a^4H^2 \int d^3 x h^{ij}h_{ij},\label{Heff}
\end{align}
which reproduces the Heisenberg equation we found in Sec.~\ref{subsecsum}.
From Eq.~\eqref{Heff}, one can read an effective graviton mass due to the coupling to radiation\footnote{The same tachyonic mass has also been discussed in \cite{Frob:2025sfq}.}:
\begin{align}
	m_{\rm eff}^2 = - \frac{2}{5}H^2,
\end{align}
where $H$ is the physical Hubble parameter.
For massive gravitons, the explicit symmetry breaking takes place even for the IR equation of motion due to the mass, rather than the gradient.
Eq.~\eqref{eomIR} with this negative effective mass squared is given by
\begin{align}
	h_{ij}'' + 2 \frac{a'}{a}h_{ij}' - \frac{2}{5}\left(\frac{a'}{a}\right)^2h_{ij}=0.\label{eomsupeff}
\end{align}
Equation~\eqref{eomsupeff} implies that large gauge symmetry is explicitly broken in the dynamics of the IR graviton modes.

Indeed, the constant shift \eqref{cshift} introduces a nonvanishing source:
\begin{align}
	h_{ij}'' + 2 \frac{a'}{a}h_{ij}' - \frac{2}{5}\left(\frac{a'}{a}\right)^2( h_{ij} -\epsilon_{ij}) =0.\label{eom22}
\end{align}
When the large gauge symmetry is broken, the IR graviton modes are no longer pure gauge, and the physical observables can depend on the constant shift $\epsilon_{ij}$.

Before the radiation-dominated era, $h_{ij}$ is free, and the large symmetry should be restored.
This requirement sets the boundary condition:
\begin{align}
	\epsilon_{ij} = h_{ij}(x_R).\label{gaugecond}
\end{align}
With this choice, the superhorizon $h_{ij}$ remains constant, eliminating the tachyonic instability on superhorizon scales.

\begin{figure}
	\includegraphics[width=1\linewidth]{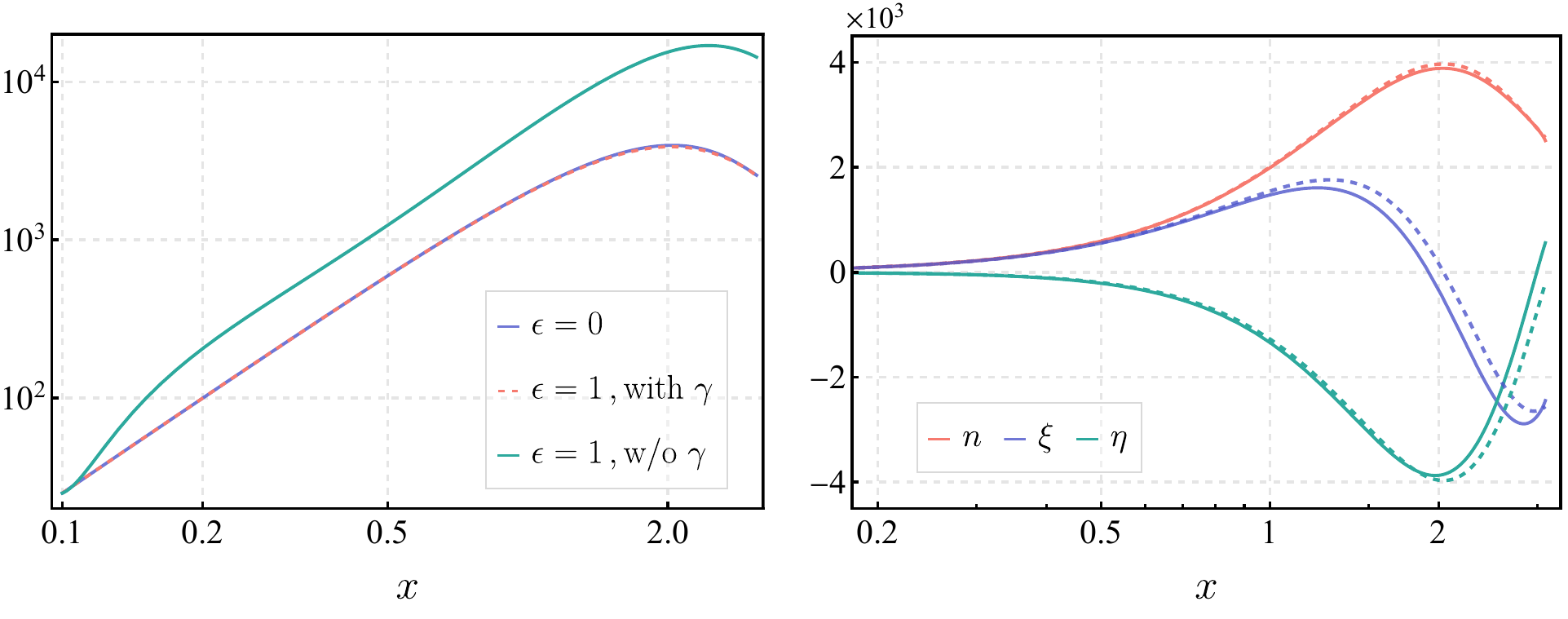}
	\caption{
	\textit{Left}: $n$ as a function of $x=k\tau$ with $x_R=0.1$ for the free theory (\textcolor[RGB]{77,85,204}{blue}), the interaction theory without regulator (\textcolor[RGB]{45,170,158}{green}), and the interaction theory with the regulator
     $\gamma\,$(\textcolor[RGB]{247,122,111}{red dashed}). The parameter $\epsilon$ is set to 1. \textit{Right}: The evolution of $n$~(\textcolor[RGB]{247,122,111}{red}), $\xi$~(\textcolor[RGB]{77,85,204}{blue}), $\eta$~(\textcolor[RGB]{45,170,158}{green}) for the free theory~(dashed) and the interaction theory with regulator $\gamma$~(solid), here parameters $\epsilon = 1$ and $x_R = 0.1$.}
	\label{Reg}
\end{figure}

Now, let us derive the effective interaction Hamiltonian after imposing the boundary condition~\eqref{gaugecond}.
Solving the linear dynamics during the radiation-dominated era gives massless evolution 
\begin{align}
	h^{(s)}_{\mathbf k}(x) = u(x-x_0) h^{(s)}_{\mathbf k}(x_0),
\end{align}
where 
\begin{align}
	u(x) \equiv \frac{\sin x}{x}\,,\label{defux}
\end{align}
and $h_{\bf_k}(x_0)$ is the amplitude at the initial time.
To be more precise, we need to take the linear solution for Eq.~\eqref{eom22}, Eq.~\eqref{defux} is enough to discuss the role as a regulator.
 In our discussion, we take $x_0=x_R$. While $u(x)$ represents the graviton mode function, its imaginary part is suppressed for $x_R=k\tau_R\to 0$.
Thus, when considering superhorizon modes at reheating, we may neglect the imaginary part.
Then, the effective mass after gauge fixing~\eqref{eom22} is recast into $\gamma m_{\rm eff}^2$ with
\begin{align}
	\gamma \equiv 1- u^{-1}.
\end{align}
Consequently, we simply replace $\alpha$ with $\gamma \alpha$ in Sec.~\ref{subsecsum}.
Since $\gamma \to 0$ in the superhorizon limit, $\gamma$ acts as an IR regulator, effectively eliminating the secular term.
After removing the secular term, the stimulated emission rate is at the percent level, consistent with calculations in the Minkowski background, as illustrated in Fig.~\ref{Reg}.

\section{Conclusions}\label{sec:conclusion}

In this paper, we considered the evolution of the graviton number operator in a thermal radiation background modeled by a scalar field. The crucial assumption was that the graviton quantum state is in a squeezed vacuum state at the initial time. This quantum state cannot be described by a classical distribution function in phase space, as the concept of particle number does not uniquely specify the state. Then, we derived the Heisenberg equation averaged over a general quantum state---the quantum state-averaged field equation of motion---rather than relying on the standard Boltzmann equation. 

The advantage of this approach is that it allows us to derive the equation of motion applicable to a nonlinear regime, similar to the Boltzmann equation. This could potentially enable partial resummation of the loop calculations in the quantum field theory approach based on the in-in formalism. Compared to previous analyses based on perturbation theory, our calculations indicate that the behavior of the system can be more accurately captured in this framework.

Depending on the initial squeezed state, the graviton number can either increase or decrease. This phenomenon, known as cosmological stimulated emission, was pointed out in Ref.~\cite{Ota:2024idm} based on perturbation theory. Like stimulated emission in QED, stronger emission is expected for more IR modes and higher temperatures. Unlike in QED, however, the soft graviton mode is unconstrained in this context. In particular, the inflationary paradigm predicts a scale-invariant primordial graviton spectrum as the initial condition.

Although we used a scalar field as a toy model of thermal radiation for simplicity, this situation is ubiquitous in the early universe, which was dominated by radiation. Therefore, our setup is quite general. We found that superhorizon gravitons are resonantly amplified in the present analysis, which contradicts the current observational status of primordial gravitational waves~\cite{Planck:2018vyg}. A potential issue with this approach is the robustness of perturbation theory in the superhorizon regime, as well as the subtlety of superhorizon gravitons.

To explore the physical picture, we first considered the same setup in the Minkowski background, where these issues do not arise. We confirmed similar stimulated emission effects in this case. 
We also confirmed that our new calculations reproduce the perturbative results for small interactions, and we observed deviation from the 1-loop perturbation theory as we increase the thermal bath temperature. However, in the parameter space where stimulated emission is significant, the backreaction to spacetime dynamics cannot be neglected. Thus, the FLRW background is essential for discussing these effects.

As a conservative scenario, we considered a radiation-dominated universe just after instantaneous reheating, focusing on the graviton modes inside the horizon at that time. These modes cannot be gauged away, and the system remains in the perturbative regime. We confirmed that stimulated emission significantly alters the initial graviton spectrum in this context.

Regarding the secular growth of superhorizon modes~\footnote{Similar secular growth is reported in Refs.~\cite{Ota:2022xni,Chen:2022dah,Ota:2022hvh} in other contexts of cosmological loop analysis.}, there may be mechanisms to suppress the effect or schemes to find a reasonable resummed value consistent with the current non-detection of primordial B-modes. A promising direction could involve considering the treatment of soft gravitons in an FLRW background. 
Free soft gravitons are considered an artifact of large gauge symmetry for the exact $k = 0$ mode. For finite-momentum modes in a decelerating universe, these gravitons eventually reenter the horizon and generate an observable signature in the sky. Hence, the role of large gauge transformations is more subtle for finite-momentum modes. Based on this perspective, we propose a regularization prescription from a symmetry-based viewpoint. 
In our interpretation, the residual large gauge symmetry in the IR effectively behaves as a global symmetry, which is \textit{explicitly} broken by an effective mass term induced by the thermal background. 
This breaking introduces a natural IR regulator, determined by a boundary condition set at reheating. 
With this regularization, the IR graviton modes remain constant on superhorizon scales, and the secular growth is eliminated.

In this paper, we did not consider the decoherence of inflationary gravitons and assumed that $\lambda \neq 0$, which may be a crucial simplification, and a more detailed analysis is left for future work.  From an engineering perspective, preparing localized radiation with high temperatures is more feasible, similar to the quark-gluon plasma generated in colliders.  While this scenario differs from the homogeneous radiation background setup assumed in our work, it remains of interest, as it may offer a practical way to realize specific thermal bath configurations that can induce strong stimulated emission without significant backreaction effects.


\appendix

\section{The formulas obtained in the in-in formalism}
\label{appA}

The graviton number and internal phase at second order in the interaction Hamiltonian are given by~\cite{Ota:2024idm}:
\begin{align}
	n_2 &=   \text{Re}\left[\zeta  (2 n_0 +1)+ 2 \sigma \lambda^*_0\right],
	\\
	\lambda_2 & =   2\zeta  \lambda_0 + \sigma (2 n_0+1) ,
\end{align}
where we define
\begin{align}
	    \zeta &\equiv 2i  \int^\tau d\tau_2 \int^\tau_{\tau_2} d\tau_1 \, u_k(\tau_2, \tau) u^*_k(\tau_1, \tau) \frac{X(\tau_1, \tau_2)}{M_{\rm pl}^2}, 
    \\
    \sigma &\equiv 2i  \int^\tau d\tau_2 \int^\tau_{\tau_2} d\tau_1 \, u_k^*(\tau_2, \tau) u^*_k(\tau_1, \tau) \frac{X(\tau_1, \tau_2)}{M_{\rm pl}^2}. 
\end{align}
The positive frequency mode function for the field operator at $\tau$, with respect to the vacuum state at $\tau_0$, is denoted by $u_k(\tau, \tau_0)$.
Unlike the calculation in the main text, here we integrate overtime twice, and the operators inside the integrand are time-ordered.
While $\lambda_2$ is not explicitly derived in the reference, its derivation closely follows that of $n_2$.

\acknowledgments

We would like to thank Toshifumi Noumi, Tomo Takahashi, Masahide Yamaguchi and Lai Han for useful discussions.
This work was supported in part by the National Natural Science Foundation of China under Grant No. 12347101 and 12403001. Y.Z. is supported by the IBS under the project code, IBS-R018-D3.

\bibliography{biblio.bib}

\begin{thebibliography}{10}

\bibitem{LIGOScientific:2016aoc}
B.~P. Abbott et~al.
\newblock {Observation of Gravitational Waves from a Binary Black Hole Merger}.
\newblock {\em Phys. Rev. Lett.}, 116(6):061102, 2016.
\newblock \href {https://arxiv.org/abs/1602.03837} {\path{arXiv:1602.03837}}, \href {https://doi.org/10.1103/PhysRevLett.116.061102} {\path{doi:10.1103/PhysRevLett.116.061102}}.

\bibitem{LIGOScientific:2017vwq}
B.~P. Abbott et~al.
\newblock {GW170817: Observation of Gravitational Waves from a Binary Neutron Star Inspiral}.
\newblock {\em Phys. Rev. Lett.}, 119(16):161101, 2017.
\newblock \href {https://arxiv.org/abs/1710.05832} {\path{arXiv:1710.05832}}, \href {https://doi.org/10.1103/PhysRevLett.119.161101} {\path{doi:10.1103/PhysRevLett.119.161101}}.

\bibitem{Planck:2018vyg}
N.~Aghanim et~al.
\newblock {Planck 2018 Results. Vi. Cosmological Parameters}.
\newblock {\em Astron. Astrophys.}, 641:A6, 2020.
\newblock [Erratum: Astron.Astrophys. 652,C \textbf{4} (2021)].
\newblock \href {https://arxiv.org/abs/1807.06209} {\path{arXiv:1807.06209}}, \href {https://doi.org/10.1051/0004-6361/201833910} {\path{doi:10.1051/0004-6361/201833910}}.

\bibitem{Ananda:2006af}
Kishore~N. Ananda, Chris Clarkson, and David Wands.
\newblock {The Cosmological gravitational wave background from primordial density perturbations}.
\newblock {\em Phys. Rev. D}, 75:123518, 2007.
\newblock \href {https://arxiv.org/abs/gr-qc/0612013} {\path{arXiv:gr-qc/0612013}}, \href {https://doi.org/10.1103/PhysRevD.75.123518} {\path{doi:10.1103/PhysRevD.75.123518}}.

\bibitem{Baumann:2007zm}
Daniel Baumann, Paul~J. Steinhardt, Keitaro Takahashi, and Kiyotomo Ichiki.
\newblock {Gravitational Wave Spectrum Induced by Primordial Scalar Perturbations}.
\newblock {\em Phys. Rev. D}, 76:084019, 2007.
\newblock \href {https://arxiv.org/abs/hep-th/0703290} {\path{arXiv:hep-th/0703290}}, \href {https://doi.org/10.1103/PhysRevD.76.084019} {\path{doi:10.1103/PhysRevD.76.084019}}.

\bibitem{Saito:2008jc}
Ryo Saito and Jun'ichi Yokoyama.
\newblock {Gravitational Wave Background as a Probe of the Primordial Black Hole Abundance}.
\newblock {\em Phys. Rev. Lett.}, 102:161101, 2009.
\newblock [Erratum: Phys.Rev.Lett. 107, 069901 (2011)].
\newblock \href {https://arxiv.org/abs/0812.4339} {\path{arXiv:0812.4339}}, \href {https://doi.org/10.1103/PhysRevLett.102.161101} {\path{doi:10.1103/PhysRevLett.102.161101}}.

\bibitem{Kohri:2018awv}
Kazunori Kohri and Takahiro Terada.
\newblock {Semianalytic calculation of gravitational wave spectrum nonlinearly induced from primordial curvature perturbations}.
\newblock {\em Phys. Rev. D}, 97(12):123532, 2018.
\newblock \href {https://arxiv.org/abs/1804.08577} {\path{arXiv:1804.08577}}, \href {https://doi.org/10.1103/PhysRevD.97.123532} {\path{doi:10.1103/PhysRevD.97.123532}}.

\bibitem{Domenech:2021ztg}
Guillem Dom\`enech.
\newblock {Scalar Induced Gravitational Waves Review}.
\newblock {\em Universe}, 7(11):398, 2021.
\newblock \href {https://arxiv.org/abs/2109.01398} {\path{arXiv:2109.01398}}, \href {https://doi.org/10.3390/universe7110398} {\path{doi:10.3390/universe7110398}}.

\bibitem{Barnaby:2011qe}
Neil Barnaby, Enrico Pajer, and Marco Peloso.
\newblock {Gauge Field Production in Axion Inflation: Consequences for Monodromy, non-Gaussianity in the CMB, and Gravitational Waves at Interferometers}.
\newblock {\em Phys. Rev. D}, 85:023525, 2012.
\newblock \href {https://arxiv.org/abs/1110.3327} {\path{arXiv:1110.3327}}, \href {https://doi.org/10.1103/PhysRevD.85.023525} {\path{doi:10.1103/PhysRevD.85.023525}}.

\bibitem{Cook:2011hg}
Jessica~L. Cook and Lorenzo Sorbo.
\newblock {Particle production during inflation and gravitational waves detectable by ground-based interferometers}.
\newblock {\em Phys. Rev. D}, 85:023534, 2012.
\newblock [Erratum: Phys.Rev.D 86, 069901 (2012)].
\newblock \href {https://arxiv.org/abs/1109.0022} {\path{arXiv:1109.0022}}, \href {https://doi.org/10.1103/PhysRevD.85.023534} {\path{doi:10.1103/PhysRevD.85.023534}}.

\bibitem{Anber:2012du}
Mohamed~M. Anber and Lorenzo Sorbo.
\newblock {Non-Gaussianities and chiral gravitational waves in natural steep inflation}.
\newblock {\em Phys. Rev. D}, 85:123537, 2012.
\newblock \href {https://arxiv.org/abs/1203.5849} {\path{arXiv:1203.5849}}, \href {https://doi.org/10.1103/PhysRevD.85.123537} {\path{doi:10.1103/PhysRevD.85.123537}}.

\bibitem{Salehian:2020dsf}
Borna Salehian, Mohammad~Ali Gorji, Shinji Mukohyama, and Hassan Firouzjahi.
\newblock {Analytic study of dark photon and gravitational wave production from axion}.
\newblock {\em JHEP}, 05:043, 2021.
\newblock \href {https://arxiv.org/abs/2007.08148} {\path{arXiv:2007.08148}}, \href {https://doi.org/10.1007/JHEP05(2021)043} {\path{doi:10.1007/JHEP05(2021)043}}.

\bibitem{Wang:2020uic}
Yi~Wang and Yuhang Zhu.
\newblock {Cosmological Collider Signatures of Massive Vectors from Non-Gaussian Gravitational Waves}.
\newblock {\em JCAP}, 04:049, 2020.
\newblock \href {https://arxiv.org/abs/2001.03879} {\path{arXiv:2001.03879}}, \href {https://doi.org/10.1088/1475-7516/2020/04/049} {\path{doi:10.1088/1475-7516/2020/04/049}}.

\bibitem{Niu:2022quw}
Xuce Niu, Moinul~Hossain Rahat, Karthik Srinivasan, and Wei Xue.
\newblock {Gravitational wave probes of massive gauge bosons at the cosmological collider}.
\newblock {\em JCAP}, 02:013, 2023.
\newblock \href {https://arxiv.org/abs/2211.14331} {\path{arXiv:2211.14331}}, \href {https://doi.org/10.1088/1475-7516/2023/02/013} {\path{doi:10.1088/1475-7516/2023/02/013}}.

\bibitem{An:2025mdb}
Haipeng An, Zhehan Qin, Zhong-Zhi Xianyu, and Borui Zhang.
\newblock {Primordial Stochastic Gravitational Waves from Massive Higher-Spin Bosons}.
\newblock 4 2025.
\newblock \href {https://arxiv.org/abs/2504.05389} {\path{arXiv:2504.05389}}.

\bibitem{Ota:2023iyh}
Atsuhisa Ota, Misao Sasaki, and Yi~Wang.
\newblock {One-Loop Thermal Radiation Exchange in Gravitational Wave Power Spectrum}.
\newblock 10 2023.
\newblock \href {https://arxiv.org/abs/2310.19071} {\path{arXiv:2310.19071}}.

\bibitem{Ota:2024idm}
Atsuhisa Ota.
\newblock {Cosmological Stimulated Emission}.
\newblock 12 2024.
\newblock \href {https://arxiv.org/abs/2412.20474} {\path{arXiv:2412.20474}}.

\bibitem{Raffelt:1992uj}
G.~Raffelt, G.~Sigl, and Leo Stodolsky.
\newblock {NonAbelian Boltzmann equation for mixing and decoherence}.
\newblock {\em Phys. Rev. Lett.}, 70:2363--2366, 1993.
\newblock [Erratum: Phys.Rev.Lett. 98, 069902 (2007)].
\newblock \href {https://arxiv.org/abs/hep-ph/9209276} {\path{arXiv:hep-ph/9209276}}, \href {https://doi.org/10.1103/PhysRevLett.70.2363} {\path{doi:10.1103/PhysRevLett.70.2363}}.

\bibitem{Sigl:1993ctk}
G.~Sigl and G.~Raffelt.
\newblock {General kinetic description of relativistic mixed neutrinos}.
\newblock {\em Nucl. Phys. B}, 406:423--451, 1993.
\newblock \href {https://doi.org/10.1016/0550-3213(93)90175-O} {\path{doi:10.1016/0550-3213(93)90175-O}}.

\bibitem{Kosowsky:1994cy}
Arthur Kosowsky.
\newblock {Cosmic Microwave Background Polarization}.
\newblock {\em Annals Phys.}, 246:49--85, 1996.
\newblock \href {https://arxiv.org/abs/astro-ph/9501045} {\path{arXiv:astro-ph/9501045}}, \href {https://doi.org/10.1006/aphy.1996.0020} {\path{doi:10.1006/aphy.1996.0020}}.

\bibitem{Alexander:2008fp}
Stephon Alexander, Joseph Ochoa, and Arthur Kosowsky.
\newblock {Generation of Circular Polarization of the Cosmic Microwave Background}.
\newblock {\em Phys. Rev. D}, 79:063524, 2009.
\newblock \href {https://arxiv.org/abs/0810.2355} {\path{arXiv:0810.2355}}, \href {https://doi.org/10.1103/PhysRevD.79.063524} {\path{doi:10.1103/PhysRevD.79.063524}}.

\bibitem{Bavarsad:2009hm}
E.~Bavarsad, M.~Haghighat, Z.~Rezaei, R.~Mohammadi, I.~Motie, and M.~Zarei.
\newblock {Generation of circular polarization of the CMB}.
\newblock {\em Phys. Rev. D}, 81:084035, 2010.
\newblock \href {https://arxiv.org/abs/0912.2993} {\path{arXiv:0912.2993}}, \href {https://doi.org/10.1103/PhysRevD.81.084035} {\path{doi:10.1103/PhysRevD.81.084035}}.

\bibitem{Mohammadi:2013dea}
Rohoollah Mohammadi.
\newblock {Evidence for cosmic neutrino background form CMB circular polarization}.
\newblock {\em Eur. Phys. J. C}, 74(10):3102, 2014.
\newblock \href {https://arxiv.org/abs/1312.2199} {\path{arXiv:1312.2199}}, \href {https://doi.org/10.1140/epjc/s10052-014-3089-7} {\path{doi:10.1140/epjc/s10052-014-3089-7}}.

\bibitem{Fidler:2017pkg}
Christian Fidler and Cyril Pitrou.
\newblock {Kinetic theory of fermions in curved spacetime}.
\newblock {\em JCAP}, 06:013, 2017.
\newblock \href {https://arxiv.org/abs/1701.08844} {\path{arXiv:1701.08844}}, \href {https://doi.org/10.1088/1475-7516/2017/06/013} {\path{doi:10.1088/1475-7516/2017/06/013}}.

\bibitem{Bartolo:2018igk}
Nicola Bartolo, Ahmad Hoseinpour, Giorgio Orlando, Sabino Matarrese, and Moslem Zarei.
\newblock {Photon-graviton scattering: A new way to detect anisotropic gravitational waves?}
\newblock {\em Phys. Rev. D}, 98(2):023518, 2018.
\newblock \href {https://arxiv.org/abs/1804.06298} {\path{arXiv:1804.06298}}, \href {https://doi.org/10.1103/PhysRevD.98.023518} {\path{doi:10.1103/PhysRevD.98.023518}}.

\bibitem{Bartolo:2019eac}
Nicola Bartolo, Ahmad Hoseinpour, Sabino Matarrese, Giorgio Orlando, and Moslem Zarei.
\newblock {CMB Circular and B-mode Polarization from New Interactions}.
\newblock {\em Phys. Rev. D}, 100(4):043516, 2019.
\newblock \href {https://arxiv.org/abs/1903.04578} {\path{arXiv:1903.04578}}, \href {https://doi.org/10.1103/PhysRevD.100.043516} {\path{doi:10.1103/PhysRevD.100.043516}}.

\bibitem{Manshouri:2020avm}
Hassan Manshouri, Ahmad Hoseinpour, and Moslem Zarei.
\newblock {Quantum Boltzmann equation for fermions: An attempt to calculate the NMR relaxation and decoherence times using quantum field theory techniques}.
\newblock {\em Phys. Rev. D}, 103(9):096020, 2021.
\newblock \href {https://arxiv.org/abs/2009.01917} {\path{arXiv:2009.01917}}, \href {https://doi.org/10.1103/PhysRevD.103.096020} {\path{doi:10.1103/PhysRevD.103.096020}}.

\bibitem{Zarei:2021dpb}
Moslem Zarei, Nicola Bartolo, Daniele Bertacca, Angelo Ricciardone, and Sabino Matarrese.
\newblock {Non-Markovian Open Quantum System Approach to the Early Universe: Damping of Gravitational Waves by Matter}.
\newblock {\em Phys. Rev. D}, 104(8):083508, 2021.
\newblock \href {https://arxiv.org/abs/2104.04836} {\path{arXiv:2104.04836}}, \href {https://doi.org/10.1103/PhysRevD.104.083508} {\path{doi:10.1103/PhysRevD.104.083508}}.

\bibitem{Sharifian:2025olk}
Mohammad Sharifian, Moslem Zarei, Nicola Bartolo, and Sabino Matarrese.
\newblock {Exploring gravitational impulse via quantum Boltzmann equation}.
\newblock 1 2025.
\newblock \href {https://arxiv.org/abs/2501.13678} {\path{arXiv:2501.13678}}.

\bibitem{Albrecht:1992kf}
Andreas Albrecht, Pedro Ferreira, Michael Joyce, and Tomislav Prokopec.
\newblock {Inflation and Squeezed Quantum States}.
\newblock {\em Phys. Rev. D}, 50:4807--4820, 1994.
\newblock \href {https://arxiv.org/abs/astro-ph/9303001} {\path{arXiv:astro-ph/9303001}}, \href {https://doi.org/10.1103/PhysRevD.50.4807} {\path{doi:10.1103/PhysRevD.50.4807}}.

\bibitem{Ghiglieri:2015nfa}
J.~Ghiglieri and M.~Laine.
\newblock {Gravitational Wave Background from Standard Model Physics: Qualitative Features}.
\newblock {\em JCAP}, 07:022, 2015.
\newblock \href {https://arxiv.org/abs/1504.02569} {\path{arXiv:1504.02569}}, \href {https://doi.org/10.1088/1475-7516/2015/07/022} {\path{doi:10.1088/1475-7516/2015/07/022}}.

\bibitem{Ghiglieri:2020mhm}
J.~Ghiglieri, G.~Jackson, M.~Laine, and Y.~Zhu.
\newblock {Gravitational Wave Background from Standard Model Physics: Complete Leading Order}.
\newblock {\em JHEP}, 07:092, 2020.
\newblock \href {https://arxiv.org/abs/2004.11392} {\path{arXiv:2004.11392}}, \href {https://doi.org/10.1007/JHEP07(2020)092} {\path{doi:10.1007/JHEP07(2020)092}}.

\bibitem{Ghiglieri:2022rfp}
Jacopo Ghiglieri, Jan Sch\"utte-Engel, and Enrico Speranza.
\newblock {Freezing-in gravitational waves}.
\newblock {\em Phys. Rev. D}, 109(2):023538, 2024.
\newblock \href {https://arxiv.org/abs/2211.16513} {\path{arXiv:2211.16513}}, \href {https://doi.org/10.1103/PhysRevD.109.023538} {\path{doi:10.1103/PhysRevD.109.023538}}.

\bibitem{Weinberg:2015QM}
Steven Weinberg.
\newblock {\em {Lectures on Quantum Mechanics}}.
\newblock 2015.
\newblock \href {https://doi.org/10.1017/CBO9781316276105} {\path{doi:10.1017/CBO9781316276105}}.

\bibitem{Frob:2025sfq}
Markus~B. Fr\"ob, Dra\v{z}en Glavan, Paolo Meda, and Ignacy Sawicki.
\newblock {One-loop correction to primordial tensor modes during radiation era}.
\newblock 4 2025.
\newblock \href {https://arxiv.org/abs/2504.02609} {\path{arXiv:2504.02609}}.

\bibitem{Emond:2018ybc}
William~T. Emond, Peter Millington, and Paul~M. Saffin.
\newblock {Boltzmann equations for preheating}.
\newblock {\em JCAP}, 09:041, 2018.
\newblock \href {https://arxiv.org/abs/1807.11726} {\path{arXiv:1807.11726}}, \href {https://doi.org/10.1088/1475-7516/2018/09/041} {\path{doi:10.1088/1475-7516/2018/09/041}}.

\bibitem{Moroi:2020bkq}
Takeo Moroi and Wen Yin.
\newblock {Particle Production from Oscillating Scalar Field and Consistency of Boltzmann Equation}.
\newblock {\em JHEP}, 03:296, 2021.
\newblock \href {https://arxiv.org/abs/2011.12285} {\path{arXiv:2011.12285}}, \href {https://doi.org/10.1007/JHEP03(2021)296} {\path{doi:10.1007/JHEP03(2021)296}}.

\bibitem{Volpe:2013uxl}
Cristina Volpe, Daavid V\"a\"an\"anen, and Catalina Espinoza.
\newblock {Extended evolution equations for neutrino propagation in astrophysical and cosmological environments}.
\newblock {\em Phys. Rev. D}, 87(11):113010, 2013.
\newblock \href {https://arxiv.org/abs/1302.2374} {\path{arXiv:1302.2374}}, \href {https://doi.org/10.1103/PhysRevD.87.113010} {\path{doi:10.1103/PhysRevD.87.113010}}.

\bibitem{Volpe:2015rla}
Cristina Volpe.
\newblock {Neutrino Quantum Kinetic Equations}.
\newblock {\em Int. J. Mod. Phys. E}, 24(09):1541009, 2015.
\newblock \href {https://arxiv.org/abs/1506.06222} {\path{arXiv:1506.06222}}, \href {https://doi.org/10.1142/S0218301315410098} {\path{doi:10.1142/S0218301315410098}}.

\bibitem{Vaananen:2013qja}
D.~V\"a\"an\"anen and C.~Volpe.
\newblock {Linearizing neutrino evolution equations including neutrino-antineutrino pairing correlations}.
\newblock {\em Phys. Rev. D}, 88:065003, 2013.
\newblock \href {https://arxiv.org/abs/1306.6372} {\path{arXiv:1306.6372}}, \href {https://doi.org/10.1103/PhysRevD.88.065003} {\path{doi:10.1103/PhysRevD.88.065003}}.

\bibitem{Serreau:2014cfa}
Julien Serreau and Cristina Volpe.
\newblock {Neutrino-antineutrino correlations in dense anisotropic media}.
\newblock {\em Phys. Rev. D}, 90(12):125040, 2014.
\newblock \href {https://arxiv.org/abs/1409.3591} {\path{arXiv:1409.3591}}, \href {https://doi.org/10.1103/PhysRevD.90.125040} {\path{doi:10.1103/PhysRevD.90.125040}}.

\bibitem{Kartavtsev:2015eva}
A.~Kartavtsev, G.~Raffelt, and H.~Vogel.
\newblock {Neutrino propagation in media: Flavor-, helicity-, and pair correlations}.
\newblock {\em Phys. Rev. D}, 91(12):125020, 2015.
\newblock \href {https://arxiv.org/abs/1504.03230} {\path{arXiv:1504.03230}}, \href {https://doi.org/10.1103/PhysRevD.91.125020} {\path{doi:10.1103/PhysRevD.91.125020}}.

\bibitem{Kanno:2021vwu}
Sugumi Kanno and Jiro Soda.
\newblock {Squeezed Quantum States of Graviton and Axion in the Universe}.
\newblock {\em Int. J. Mod. Phys. D}, 31(13):2250098, 2022.
\newblock \href {https://arxiv.org/abs/2112.14496} {\path{arXiv:2112.14496}}, \href {https://doi.org/10.1142/S0218271822500985} {\path{doi:10.1142/S0218271822500985}}.

\bibitem{Weinberg:2005vy}
Steven Weinberg.
\newblock {Quantum Contributions to Cosmological Correlations}.
\newblock {\em Phys. Rev. D}, 72:043514, 2005.
\newblock \href {https://arxiv.org/abs/hep-th/0506236} {\path{arXiv:hep-th/0506236}}, \href {https://doi.org/10.1103/PhysRevD.72.043514} {\path{doi:10.1103/PhysRevD.72.043514}}.

\bibitem{Weinberg:2003sw}
Steven Weinberg.
\newblock {Adiabatic Modes in Cosmology}.
\newblock {\em Phys. Rev. D}, 67:123504, 2003.
\newblock \href {https://arxiv.org/abs/astro-ph/0302326} {\path{arXiv:astro-ph/0302326}}, \href {https://doi.org/10.1103/PhysRevD.67.123504} {\path{doi:10.1103/PhysRevD.67.123504}}.

\bibitem{Sloth:2025nan}
Martin~S. Sloth.
\newblock {Soft Gravitons as Goldstone Modes of Spontaneously Broken Asymptotic Symmetries in De Sitter}.
\newblock 2 2025.
\newblock \href {https://arxiv.org/abs/2502.03520} {\path{arXiv:2502.03520}}.

\bibitem{Pajer:2017hmb}
Enrico Pajer and Sadra Jazayeri.
\newblock {Systematics of Adiabatic Modes: Flat Universes}.
\newblock {\em JCAP}, 03:013, 2018.
\newblock \href {https://arxiv.org/abs/1710.02177} {\path{arXiv:1710.02177}}, \href {https://doi.org/10.1088/1475-7516/2018/03/013} {\path{doi:10.1088/1475-7516/2018/03/013}}.

\bibitem{Ota:2022xni}
Atsuhisa Ota, Misao Sasaki, and Yi~Wang.
\newblock {One-Loop Tensor Power Spectrum from an Excited Scalar Field during Inflation}.
\newblock {\em Phys. Rev. D}, 108(4):043542, 2023.
\newblock \href {https://arxiv.org/abs/2211.12766} {\path{arXiv:2211.12766}}, \href {https://doi.org/10.1103/PhysRevD.108.043542} {\path{doi:10.1103/PhysRevD.108.043542}}.

\bibitem{Chen:2022dah}
Chao Chen, Atsuhisa Ota, Hui-Yu Zhu, and Yuhang Zhu.
\newblock {Missing One-Loop Contributions in Secondary Gravitational Waves}.
\newblock {\em Phys. Rev. D}, 107(8):083518, 2023.
\newblock \href {https://arxiv.org/abs/2210.17176} {\path{arXiv:2210.17176}}, \href {https://doi.org/10.1103/PhysRevD.107.083518} {\path{doi:10.1103/PhysRevD.107.083518}}.

\bibitem{Ota:2022hvh}
Atsuhisa Ota, Misao Sasaki, and Yi~Wang.
\newblock {Scale-Invariant Enhancement of Gravitational Waves during Inflation}.
\newblock {\em Mod. Phys. Lett. A}, 38(12n13):2350063, 2023.
\newblock \href {https://arxiv.org/abs/2209.02272} {\path{arXiv:2209.02272}}, \href {https://doi.org/10.1142/S0217732323500633} {\path{doi:10.1142/S0217732323500633}}.

\end{thebibliography}
\bibliographystyle{unsrturl}

\end{document}